\documentclass[conference,10pt]{IEEEtran}
\IEEEoverridecommandlockouts   
\usepackage[utf8]{inputenc}
\usepackage{newtxtext,newtxmath}
\usepackage[dvipsnames]{xcolor}
\usepackage[T1]{fontenc}
\usepackage{array}
\usepackage{url}
\usepackage{balance}
\usepackage{graphicx}
\usepackage{algorithmic}
\usepackage{hyperref}
\usepackage{microtype}
\usepackage{enumitem}

\ifCLASSOPTIONcompsoc
  \usepackage[caption=false, font=normalsize, labelfont=sf, textfont=sf]{subfig}
  \usepackage[nocompress]{cite}
\else
  \usepackage[caption=false, font=footnotesize]{subfig}
  \usepackage{cite}
\fi

\usepackage[acronym,nohypertypes=acronym]{glossaries}
\usepackage[capitalise]{cleveref}

\setacronymstyle{long-short}

\IEEEtriggercmd{\balance}
\IEEEtriggeratref{50}

\newcommand{\papertitle}{Talking After Lights Out: An Ad Hoc Network for Electric Grid Recovery}

\hypersetup{
  colorlinks=true,
  linkcolor=black,
  urlcolor=blue,
  citecolor=black,
  pdftitle={\papertitle}
}

\newacronym{DARPA}{DARPA}{Defense Advanced Research Projects Agency}
\newacronym{SCADA}{SCADA}{supervisory control and data acquisition}
\newacronym{RADICS}{RADICS}{Rapid Attack Detection, Isolation, and Characterization Systems}
\newacronym{PhoenixSEN}{PhoenixSEN}{Phoenix Secure Emergency Network}
\newacronym{NETS}{NETS}{national electricity transmission system}
\newacronym{RTO}{RTO}{regional transmission organization}
\newacronym{ISP}{ISP}{internet service provider}
\newacronym{NIST}{NIST}{National Institute of Standards and Technology}
\newacronym{IT}{IT}{information technology}
\newacronym{VoIP}{VoIP}{Voice over IP}
\newacronym{OS}{OS}{operating system}
\newacronym{IP}{IP}{Internet Protocol}
\newacronym{ICMP}{ICMP}{Internet Control Message Protocol}
\newacronym{TCP}{TCP}{Transmission Control Protocol}
\newacronym{SNMP}{SNMP}{Simple Network Monitoring Protocol}
\newacronym{GPS}{GPS}{Global Positioning System}
\newacronym{ISO}{ISO}{independent system operator}
\newacronym{CPS}{CPS}{cyber-physical system}
\newacronym{FERC}{FERC}{Federal Energy Regulation Commission}
\newacronym{NERC}{NERC}{North American Electric Reliability Corporation}
\newacronym{ICS}{ICS}{industrial control system}
\newacronym{EMS}{EMS}{energy management system}
\newacronym{OASIS}{OASIS}{Open Access Same-Time Information System}
\newacronym{HTTP}{HTTP}{Hypertext Transfer Protocol}
\newacronym{API}{API}{application programming interface}
\newacronym{WAN}{WAN}{wide area network}
\newacronym{MPLS}{MPLS}{Multiprotocol Label Switching}
\newacronym{PSTN}{PSTN}{public switched telephone network}
\newacronym{FAN}{FAN}{field area network}
\newacronym{MTU}{MTU}{master terminal unit}
\newacronym{RTU}{RTU}{remote terminal unit}
\newacronym{PDC}{PDC}{phasor data concentrator}
\newacronym{PMU}{PMU}{phasor measurement unit}
\newacronym{NUC}{NUC}{Next Unit of Computing}
\newacronym{OLSR}{OLSR}{Optimized Link State Routing Protocol}
\newacronym{CC}{CC}{control center}
\newacronym{HF}{HF}{high frequency}
\newacronym{VLAN}{VLAN}{virtual LAN}
\newacronym{LAN}{LAN}{local area network}
\newacronym{DNS}{DNS}{Domain Name System}
\newacronym{DHCP}{DHCP}{Dynamic Host Configuration Protocol}
\newacronym{NTP}{NTP}{Network Time Protocol}
\newacronym{IDS}{IDS}{intrusion detection system}
\newacronym{MANET}{MANET}{mobile ad hoc network}
\newacronym{USB}{USB}{Universal Serial Bus}
\newacronym{VxLAN}{VxLAN}{Virtual Extensible LAN}
\newacronym{UDP}{UDP}{User Datagram Protocol}
\newacronym{VNI}{VNI}{virtual network identifier}
\newacronym{SIP}{SIP}{Session Initiation Protocol}
\newacronym{TLS}{TLS}{Transport Layer Security}
\newacronym{DNS-SD}{DNS-SD}{DNS Service Discovery}
\newacronym{mDNS}{mDNS}{multicast DNS}
\newacronym{BMF}{BMF}{Basic Multicast Forwarding}
\newacronym{URI}{URI}{Uniform Resource Identifier}
\newacronym{WebRTC}{WebRTC}{Web Real-Time Communication}
\newacronym{DECT}{DECT}{digital enhanced cordless communications}
\newacronym{UI}{UI}{user interface}
\newacronym{ARP}{ARP}{Address Resolution Protocol}
\newacronym{MITM}{MITM}{man-in-the-middle}
\newacronym{PNAC}{PNAC}{Post-based Network Access Control}
\newacronym{EAPoL}{EAPoL}{Extensible Authentication Protocol over LAN}
\newacronym{RADIUS}{RADIUS}{Remote Authentication Dial-In User Service}
\newacronym{PTT}{PTT}{push-to-talk}
\newacronym{PLC}{PLC}{power line communication}
\newacronym{RoHC}{RoHC}{Robust Header Compression}
\newacronym{SBC}{SBC}{session border controller}
\newacronym{DNP3}{DNP3}{Distributed Network Protocol 3}
\newacronym{ICCN}{ICCN}{Inter-Control Center Communications Protocol}
\newacronym{IEC}{IEC}{International Electrotechnical Commission}
\newacronym{DIY}{DIY}{do it yourself}
\newacronym{US}{U.S.}{United States}

\begin{document}
\title{\papertitle}

\author{
  \IEEEauthorblockN{
    \href{mailto:janakj@cs.columbia.edu}{\color{black}Jan Janak}\IEEEauthorrefmark{1},
    \href{mailto:dchee@perspectalabs.com}{\color{black}Dana Chee}\IEEEauthorrefmark{2},
    \href{mailto:hema.retty@baesystems.com}{\color{black}Hema Retty}\IEEEauthorrefmark{3},
    \href{mailto:ab4659@columbia.edu}{\color{black}Artiom Baloian}\IEEEauthorrefmark{1},
    \href{mailto:hgs@cs.columbia.edu}{\color{black}Henning Schulzrinne}\IEEEauthorrefmark{1}
  }
  \IEEEauthorblockA{%
    \IEEEauthorrefmark{1}Department of Computer Science, Columbia University, USA
  }
  \IEEEauthorblockA{%
    \IEEEauthorrefmark{2}Perspecta Labs, USA
  }
  \IEEEauthorblockA{%
    \IEEEauthorrefmark{3}FAST Labs, BAE Systems, USA
  }%
  Email:
    \href{mailto:janakj@cs.columbia.edu}{\color{black}janakj@cs.columbia.edu}, \href{mailto:dchee@perspectalabs.com}{\color{black}dchee@perspectalabs.com}, \href{mailto:hema.retty@baesystems.com}{\color{black}hema.retty@baesystems.com},\\%
    \href{mailto:ab4659@columbia.edu}{\color{black}ab4659@columbia.edu},
    \href{mailto:hgs@cs.columbia.edu}{\color{black}hgs@cs.columbia.edu}%
  \thanks{%
    This research was developed with funding from the \gls{DARPA}. The views and conclusions contained in this document are those of the authors and should not be interpreted as representing the official policies, either expressed or implied, of the \glsdesc{DARPA} or the U.S. government. Distribution statement A. Distribution approved for public release, distribution unlimited. Not export controlled per ES-FL-020821-0013.
  }
}

\maketitle

\glsresetall
\begin{abstract}
When the electric grid in a region suffers a major outage, e.g., after a catastrophic cyber attack, a ``black start'' may be required, where the grid is slowly restarted, carefully and incrementally adding generating capacity and demand. To ensure safe and effective black start, the grid control center has to be able to communicate with field personnel and with \gls{SCADA} systems. Voice and text communication are particularly critical. As part of the \gls{DARPA} \gls{RADICS} program, we designed, tested and evaluated a self-configuring mesh network architecture and prototype called the \gls{PhoenixSEN}. \gls{PhoenixSEN} is designed as a drop-in replacement for primary communication networks, combines existing and new technologies, can work with a variety of link-layer protocols, emphasizes manageability and auto-configuration, and provides a core set of services and applications for coordination of people and devices including voice, text, and \gls{SCADA} communication. The \gls{PhoenixSEN} prototype was evaluated in the field through a series of \gls{DARPA}[-led] exercises. The same system is also likely to support coordination of recovery efforts after large-scale natural disasters.
\end{abstract}

\begin{IEEEkeywords}
Ad hoc networks, network architecture, network security.
\end{IEEEkeywords}

\glsresetall
\section{Introduction}\label{sec:introduction}


%

Most electric power outages are locally-contained and recovery can rely on the public or utility-owned communications infrastructure to coordinate restoration and energizing parts of the electric grid. Large-scale electric power outages, a.k.a blackouts, are rare but do happen~\cite{argentina2019,uk2019}. Recovering from a large-scale outage typically follows a special procedure known as black start. ``A total or partial shutdown of the \gls{NETS} is an unlikely event. However, if it happens, we are obliged to make sure there are contingency arrangements in place to ensure electricity supplies can be restored in a timely and orderly way. Black start is a procedure to recover from such a shutdown.''~\cite{NationalGridESO}

In the \gls{US}, the black start procedure is usually managed by \glspl{RTO} that coordinate several electric utilities. For example, PJM, a large \gls{RTO}, describes its black start operation as follows: ``Black Start capability is necessary to restore the PJM transmission system following a blackout. Black Start Service shall enable PJM, in collaboration with the Transmission Owners, to designate specific generators whose location and capabilities are required to re-energize the transmission system.''~\cite{PJM12}

Even if sufficient black start generating capability is available, a successful black start requires coordination of electricity supply and demand, typically by incrementally adding both generating capacity and load. Such coordination usually takes place either via phone calls to substation personnel, or via real-time control of \gls{SCADA} devices. Both cases require network connectivity. Grid operators often rely on \glspl{ISP} for network services~\cite{nist-smartgrid-roadmap}. If the \glspl{ISP} are also impaired by the blackout, network connectivity may be difficult to guarantee.

If the blackout is caused by a network-based cyber attack, the attacker may also attempt to actively thwart or delay power restoration, making a bad situation worse. The \gls{DARPA} has recognized the danger network-based cyber attacks represent for the \gls{US} critical power grid infrastructure and launched the \gls{RADICS} program~\cite{radics}. The goal of the program is to create a set of tools that will aid the power distribution industry in recovering from a hypothetical large-scale blackout triggered by a network-based cyber attack.

In this paper, we present the design, prototype implementation, and experimental evaluation of the \gls{PhoenixSEN} designed by BAE Systems and Columbia University as part of the \gls{DARPA} \gls{RADICS} tool set. \gls{PhoenixSEN} consists of a hybrid, isolated, self-forming network and services specifically designed to enable the coordination of power restoration amidst an ongoing network-based cyber attack. It combines existing and new technologies, can work with a variety of link-layer protocols, and provides applications for rapid coordination of people and devices. The network is designed as a drop-in replacement for primary communication networks that are likely to be severely impaired during a large-scale blackout.

We begin by discussing the motivation and the problem being addressed by our work (and the \gls{DARPA} \gls{RADICS} program in general) in \cref{sec:motivation-problem-statement}. \cref{sec:electrical-grid-model} presents a simplified model of the \gls{US} electrical grid with an emphasis on networking infrastructure. We then discuss the general architecture and features of the Phoenix network and node in \cref{sec:phoenix-sen}. The subsequent sections describe various building blocks, applications, and services in detail: naming and service discovery (\cref{sec:naming-discovery}), voice and chat support (\cref{sec:voice-chat}), network monitoring (\cref{sec:netmon}), and insider attack mitigation (\cref{sec:ethershield}). We conclude in \cref{sec:conclusion} and discuss potential future work and applicability of \gls{PhoenixSEN} to other scenarios, e.g., natural disaster response.

\section{Motivation \& Problem Statement}\label{sec:motivation-problem-statement}

Our work is primarily motivated by the need of the power distribution industry for backup network infrastructure that could be used to recover from a large-scale blackout caused by a network-based cyber attack~\cite{idaho}. We assume that the grid control \gls{SCADA} devices, as well as the network infrastructure itself maybe be compromised and could act maliciously. Therefore, the backup infrastructure should provide a means to reconnect healthy devices and keep those isolated from potentially compromised or malicious devices. This would allow for an incremental approach where devices are connected to the temporary (isolated) network only after they have been inspected and deemed healthy.

Like most complex networked systems today, power grid infrastructure relies, at least partially, on networks managed by external \glspl{ISP} for remote command, control, and coordination of both machines (\gls{SCADA}) and operators (voice). It is likely that the network infrastructure itself will be severely affected in case of a large-scale blackout. This presents an interesting ``chicken or the egg'' dilemma. The network needs power to connect the grid, but the grid cannot operate without the network. Clearly, there needs to be infrastructure in place that will allow the grid to be temporarily self-sufficient, at least during the initial restart phase when the grid is not yet fully operational. We envision a mostly self-configuring network infrastructure that can take advantage of various existing link layer technologies and can be deployed to geographically dispersed sites used by the power grid infrastructure after the blackout. We assume the personnel setting up the infrastructure will have technical background, but not necessarily in computer or network engineering. The network would provide the minimum set of services and bandwidth necessary to black-start the power grid in a secure manner.

A recent series of large scale blackouts illustrates that such events are not merely a hypothetical possibility. The 2019 blackout in Argentina, Uruguay, and Paraguay left 48 million people without power~\cite{argentina2019}. For comparison, the Boston-Washington metropolitan area has 50 million inhabitants. In 2019, a large-scale power outage in England affected more than a million homes and severely disrupted the public transit system~\cite{uk2019,uk2019b}.

Network-based cyber attacks on critical power grid infrastructure can potentially have even more disastrous and long-lasting consequences, leaving tens of millions of people in large metropolitan areas without power for extended periods of time. An ongoing cyber attack on power grid systems may attempt to thwart any restart attempts, leaving a large number of people without power for days or weeks. Depending on the state of the power grid infrastructure, a full black-start recovery after a cyber attack may also take a long time, e.g., days or weeks, depending on the sophistication of the attack.

While some of the critical grid infrastructure may be temporarily powered by on-site backup diesel generators, the situation will get progressively worse as those generators begin malfunctioning or start running out of fuel. Communication networks stop working, potentially bringing down other critical services such as the \gls{GPS}. Water and gas systems, hospitals, nursing homes, and waste treatment facilities will soon begin shutting down, transportation infrastructure can be severely affected. The situation gets progressively worse as more critical services shut down~\cite{nist-disaster}.

In a large-scale catastrophic scenario like this, with a substantial and prolonged disruption of electric power, time is of the essence. The operation of the electrical grid must be restored within a few days to prevent other critical facilities from shutting down. A full black-start recovery of the electric grid requires communication and coordination. However, communication networks are unlikely to remain operational after a substantial power disruption event.

While the main use case for the work presented in this paper is black-start recovery of the power grid, a similar architecture could also be used to create temporary isolated networks for emergency communications in various disaster relief scenarios, e.g., the hurricanes that ravaged Puerto Rico in 2017~\cite{fcc-hurricane-impact}. Similar technology is being used by some disaster relief organizations such as the Red Cross~\cite{redcross}.




\section{Architectural Model of U.S. Electrical Grid}\label{sec:electrical-grid-model}

\begin{figure*}
  \subfloat[Distributed grid model where electricity market participants use shared transmission infrastructure coordinated by a regional transmission operator (RTO) or independent system operator (ISO).]
  {\includegraphics[width=0.427\linewidth]{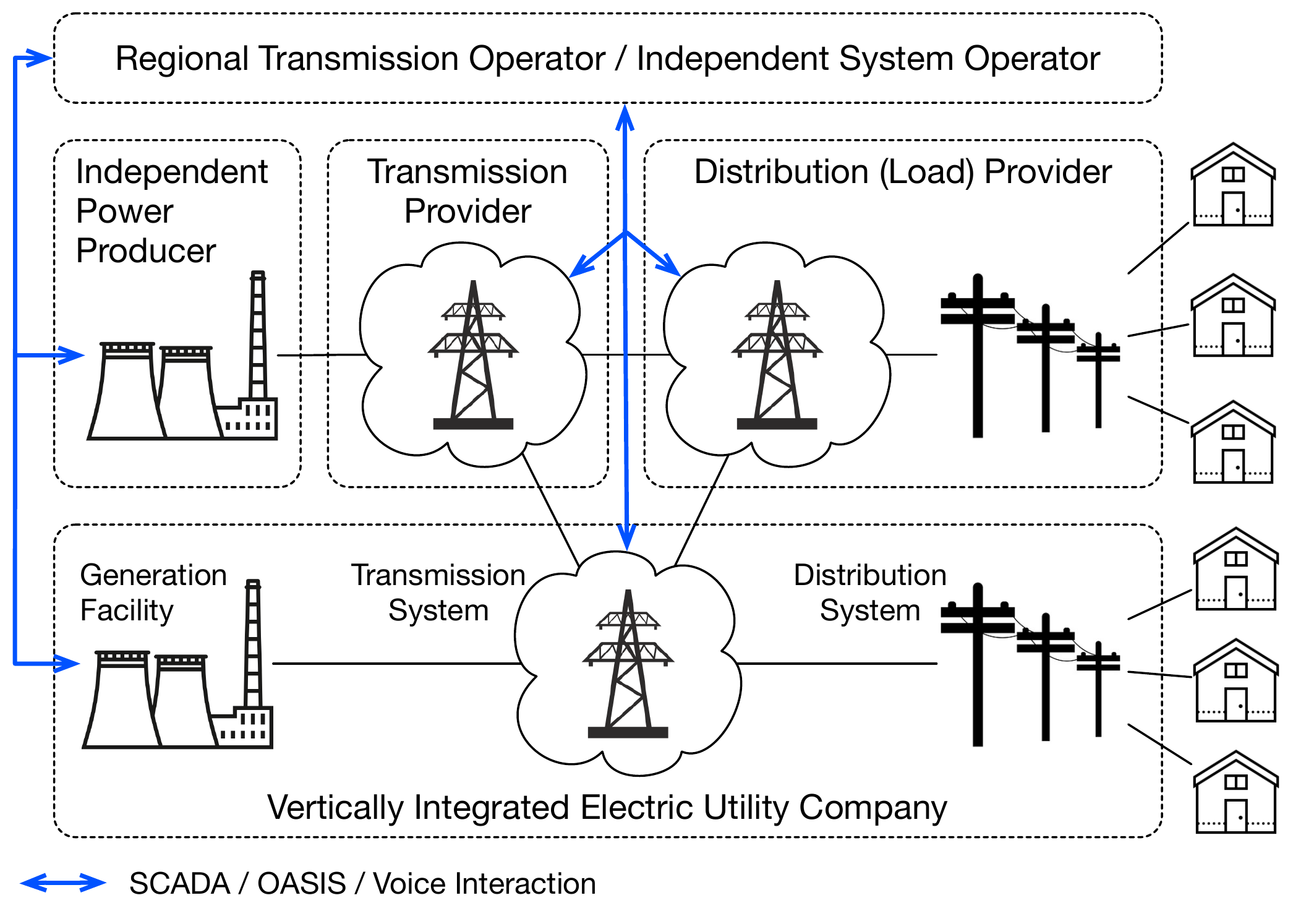}}
  \hspace*{\fill}
  \subfloat[A detailed overview of the communications infrastructure required to keep the grid operational and reliable. The diagram shows a simplified architectural model. The existing grid exhibits considerable variety across geographic and political boundaries.]
  {\includegraphics[width=0.56\linewidth]{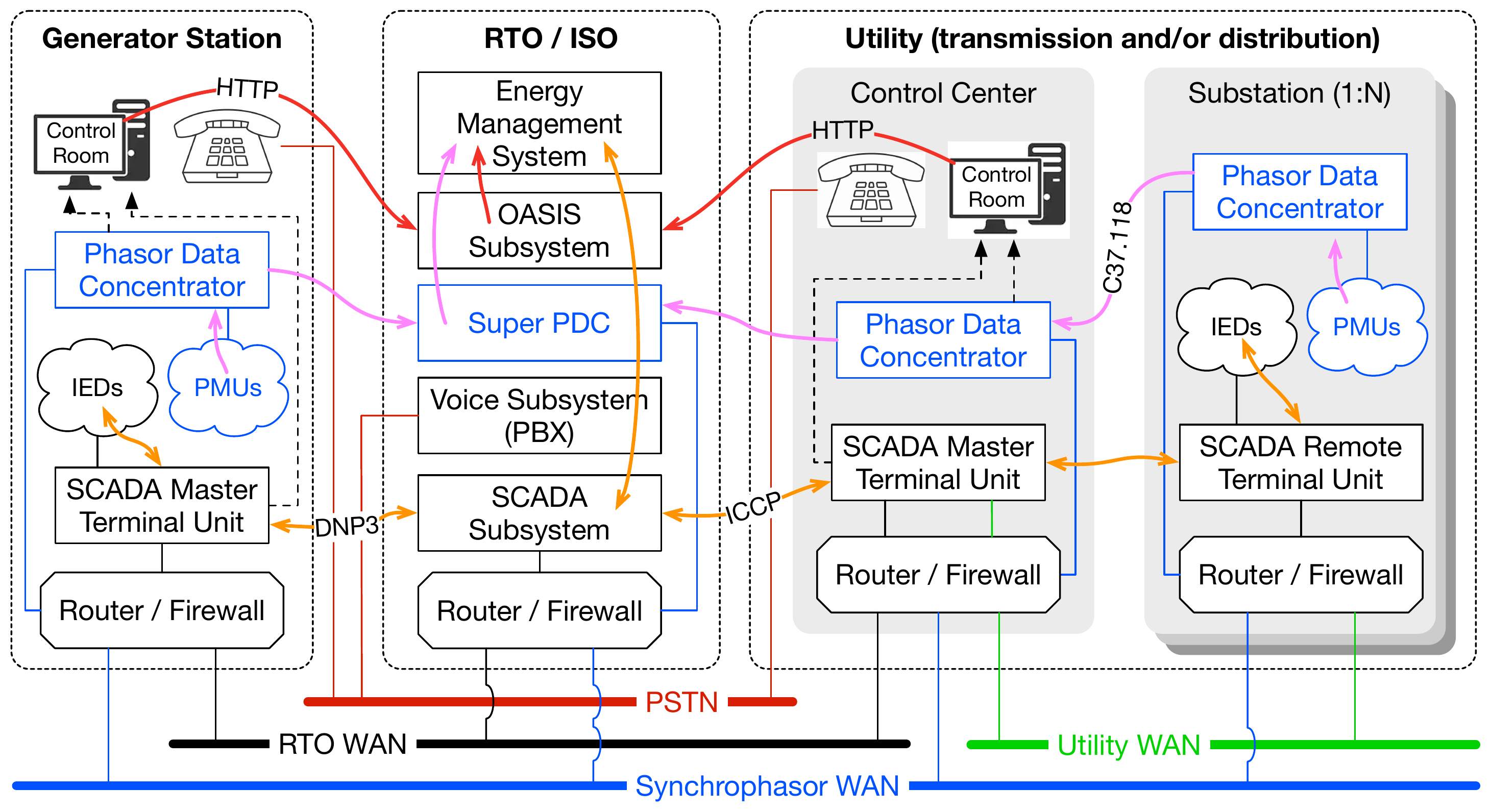}}
  \caption{The U.S. electrical grid evolves towards a more distributed model which requires an increasing amount of coordination, which in turn requires extensive data communications infrastructure. Modern-day grid infrastructure resembles a geographically dispersed cyber-physical system (CPS), where a computer program (EMS) manages the flow of electric power through the system via real-time \gls{SCADA}[-based] remote instrumentation of physical processes.}\label{fig:grid-architecture}
\end{figure*}

The \gls{US} electrical grid is a complex, heterogeneous, geographically dispersed system that combines physical infrastructure for producing and delivering electric power with computer-based monitoring, management, and control. As essential infrastructure that has been evolving for more than a century and is subject to extensive government regulation, the grid has seen incremental upgrades and organic growth, resulting in considerable variability across geographical and political boundaries. The grid's architecture is moving from a model with a small number of vertically-integrated\footnote{A vertically-integrated utility controls all stages of electric power supply chain, from generation to distribution among consumers.} utility monopolies towards an interconnected model with a multitude of utility and non-utility companies coordinating to use shared transmission infrastructure. In the interconnected model, networking infrastructure plays an increasingly important role and is critical for reliable grid operation.

In the \gls{US}, hundreds of companies participate in the production and transmission of electric power. To ensure reliable grid operation, the \gls{FERC} has designated the \gls{NERC} to develop and enforce operational standards. Regional system operators coordinate the generation, transmission, and distribution of electric power. In some regions the system operator is affiliated with a particular utility company. In other regions the system operator is an independent entity known as the \gls{RTO} that coordinates multiple utilities. Some regions have an \gls{ISO} instead with a similar role. The differences are subtle and beyond the scope of this paper. The \gls{RTO}/\gls{ISO} operates a wholesale electricity market, guarantees non-discriminatory access to shared grid infrastructure, and ensures reliable grid operation and compliance with \gls{NERC} standards. The actual electric power generation, distribution, metering, and billing is provided by utility and non-utility companies coordinated by the \gls{RTO}/\gls{ISO}.

The major elements of an electric grid are the devices that produce and transmit electric power, \gls{IT}, \glspl{ICS}, and the underlying network infrastructure. Since electric power is generated and consumed almost instantaneously, the grid must be coordinated to match power generation with demand in real-time. \figurename~\ref{fig:grid-architecture} provides a simplified architectural model of the \gls{US} grid with a focus on the communications infrastructure.

\gls{NERC} operational standards provide high-level guidance primarily aimed at ensuring reliable grid operation. The actual implementation details of power grid systems are left to the \glspl{RTO}/\glspl{ISO} and utilities. As a result, existing grid systems are heterogeneous and often use a multitude of devices that communicate with mutually incompatible protocols. Early substation automation devices communicated using proprietary protocols over industry-specific buses or serial links. Modern-day substation systems tend to use standardized process automation protocols and reuse existing wired and wireless network technologies.

The flow of electric power through the grid is managed by an \gls{EMS} program. The primary purpose of the \gls{EMS} to keep the grid operational and reliable in response to varying conditions such as the available generator pool, transmission capacity, and instantaneous load. Ideally, a single instance of the \gls{EMS} with the ability to remotely control critical grid devices would be provided by the \gls{RTO}/\gls{ISO} for the entire region. In practice, individual transmission operators typically run their own \gls{EMS} to monitor and protect their assets.

The \gls{EMS} obtains the data about available generation resources and transmission capacity for its scheduling and planning algorithms from the \gls{OASIS}, a standardized web-based management system~\cite{ferc-889} that serves as an interface between electricity market participants, transmission providers, and the \gls{RTO}/\gls{ISO}. \gls{FERC} requires that each \gls{RTO}/\gls{ISO} must provide an \gls{OASIS} node. Clients typically interact with the \gls{OASIS} system by invoking \gls{HTTP} \glspl{API} over the internet.

The operators of infrastructure deemed critical for the reliability of the grid by the \gls{RTO}/\gls{ISO} are required to provide the \gls{RTO}/\gls{ISO} with remote access to selected components for monitoring and control purposes. This is accomplished by integrating the \gls{RTO}/\gls{ISO}['s] and the operator's \gls{SCADA} and synchrophasor subsystems over a redundant \gls{WAN} provided for this purpose by the \gls{RTO}/\gls{ISO}. The data obtained from these subsystems is used by the \gls{EMS} to build a global view of the state of the grid. Furthermore, the \gls{EMS} can use \gls{SCADA} to remotely control grid devices in the field, e.g., circuit breakers. Not all grid participants need to be \gls{SCADA}[-capable] and connected to the \gls{RTO}/\gls{ISO} \gls{WAN}. Smaller entities sometimes rely on the internet for all communication with the \gls{RTO}/\gls{ISO}.

\begin{figure*}
  \subfloat[A Phoenix node at each substation connects SCADA and backend devices to a virtual network spanning all substations of the utility. Per-utility virtual networks share common physical infrastructure but are isolated from one another. Each Phoenix node helps route packets for other utilities. A dedicated forensic access port is provided on each node.]{%
    \includegraphics[width=0.57\linewidth]{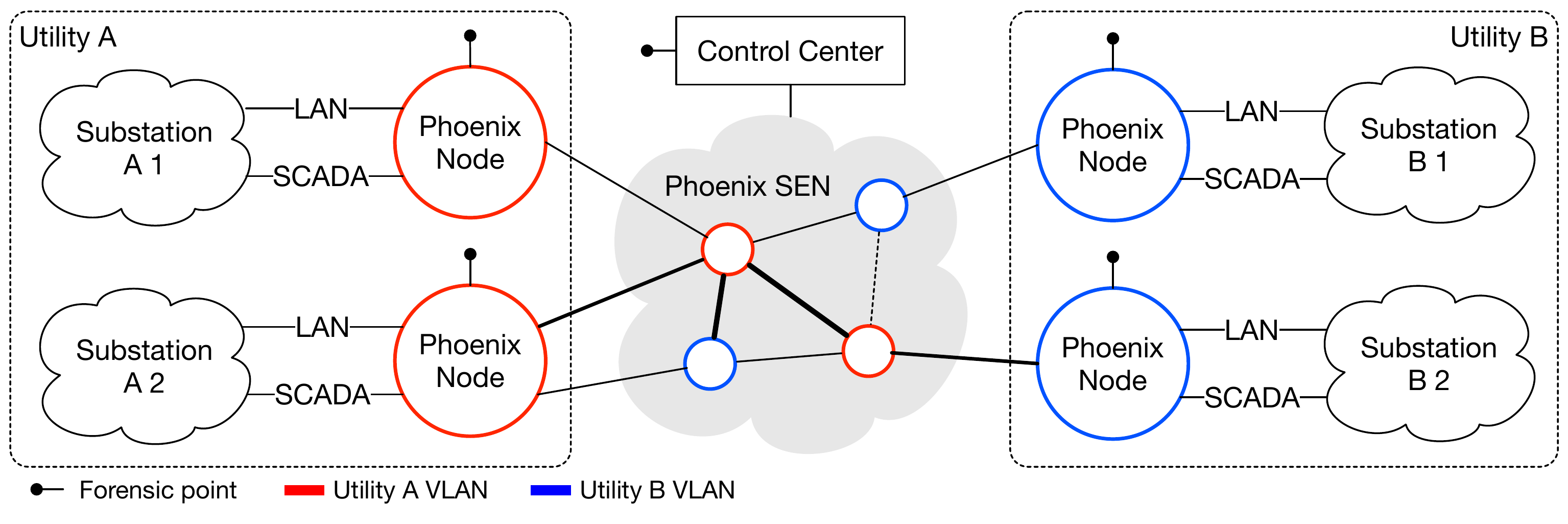}%
    \label{fig:net-arch}%
  }
  \hspace*{\fill}
  \subfloat[The Phoenix node consists of an Intel NUC with a uniform software installation and peripherals in a weather resistant enclosure. One-time deployment configuration is performed via a memory card (distributed separately) or the included smartphone.]{%
    \includegraphics[width=0.41\linewidth]{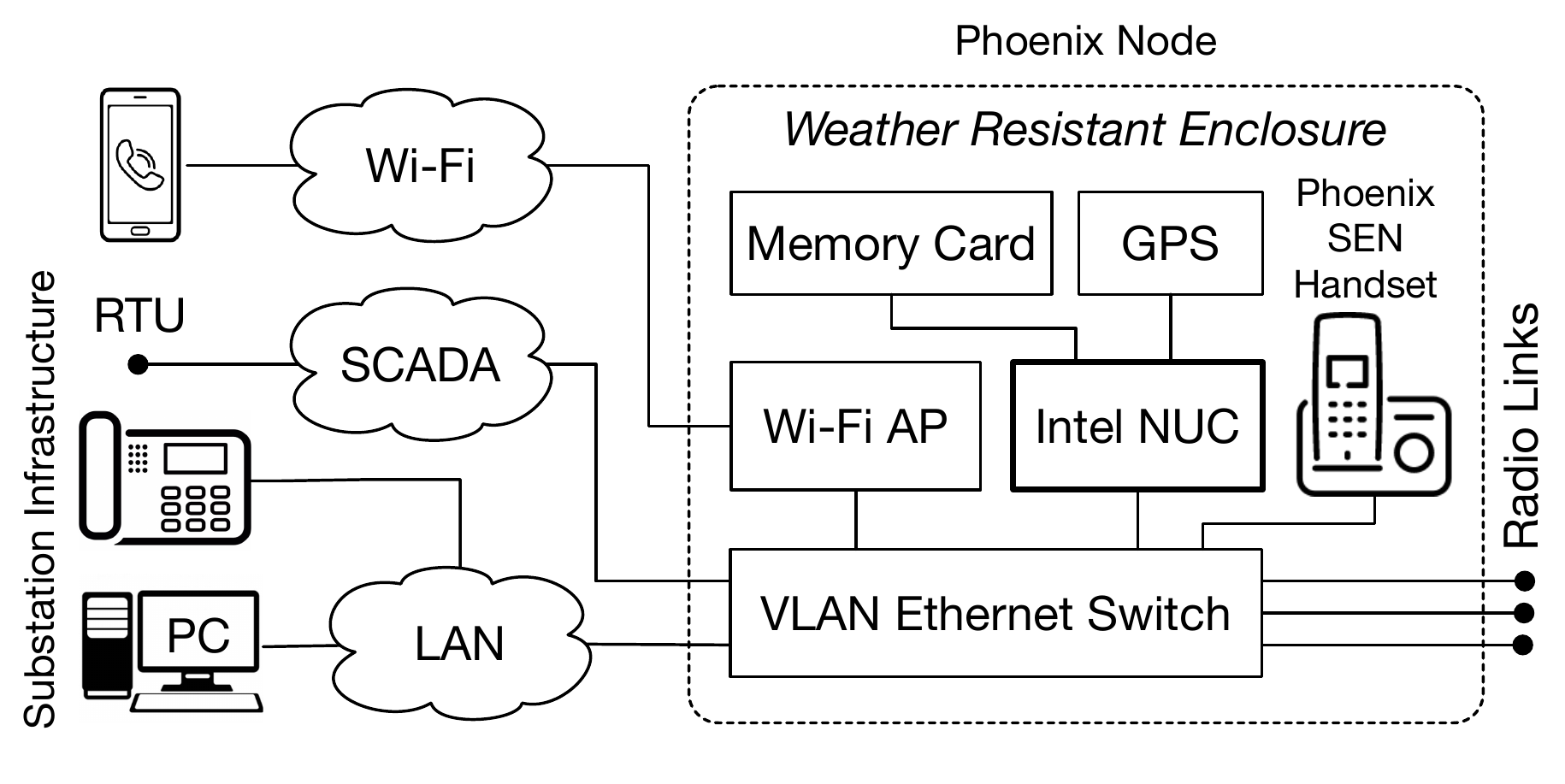}%
    \label{fig:node-arch}%
  }
  \caption{PhoenixSEN is a drop-in replacement for ISP networks used by utilities. The OLSR-based network consists of uniform Phoenix nodes deployed at substations and interconnected by a variety of links (long, short, fast, slow). Each utility is provided with an isolated virtual network spanning all its substations.}\label{fig:sen-arch}
\end{figure*}

From the previous paragraphs it follows that many different types of data networks are involved in managing the flow of electricity through the grid. The \gls{RTO}/\gls{ISO} operates a redundant \gls{WAN} used to connect to the control centers of grid infrastructure providers critical for overall system reliability. The typical \gls{RTO}/\gls{ISO} \gls{WAN} is a redundant \gls{MPLS} network based on links leased from several external \glspl{ISP}. The synchrophasor subsystem, if deemed critical for future grid systems, will most likely use a dedicated \gls{WAN} with stricter latency and bandwidth guarantees.

The use of the \gls{PSTN} for human-to-human voice communications between the \gls{RTO}/\gls{ISO} and utility personnel is required by \gls{NERC}. Despite an overall increase in remote instrumentation capabilities across the electrical grid, human-to-human voice communication remains the most important communication modality in emergency situations, e.g., after a blackout. To meet \gls{NERC}['s] strict reliability requirements, the phone system typically uses cellular or satellite phones as backup.

Each utility also operates a dedicated \gls{WAN} spanning its (sometimes large) service area that connects the utility's control center with all substations. The typical utility \gls{WAN} is \gls{IP}[-based] and uses a combination of public (\gls{ISP}[-owned]) and non-public (utility-owned) network infrastructure. A substation with connected devices (e.g., \gls{SCADA}) must also provide a \gls{FAN}. The \gls{FAN} connects substation automation devices as well as any remote devices (metering, data collection) within the substation's service area, e.g., a neighborhood. Due to the large variety in deployed automation devices, the \gls{FAN} is perhaps the most heterogeneous network type and is typically based on a combination of wired and wireless technologies. The public internet (not pictured) is typically used for other communication, e.g., to access the \gls{RTO}/\gls{ISO}['s] \gls{OASIS} portal, or to transfer metering or billing information between the utility and its customers.

\gls{SCADA} interactions between the \gls{RTO}/\gls{ISO} and utility control centers typically use standardized protocols such as the \gls{DNP3}~\cite{DNP3}, the \gls{ICCN}~\cite{ICCN}, or IEC 61850~\cite{iec-61850} carried over TCP/IP. The \gls{SCADA} subsystem is a hierarchically organized system where the \gls{RTO}/\gls{ISO}['s] \gls{MTU} communicates with the \glspl{MTU} at utility control centers, which in turn communicate with \glspl{RTU} deployed at substations. The synchrophasor subsystem is based on a hierarchy of \glspl{PDC} that aggregate and process IEEE C37.118~\cite{ieee-c37} data streams coming from \glspl{PMU} deployed across the grid.

\section{Phoenix Secure Emergency Network}\label{sec:phoenix-sen}

The \gls{PhoenixSEN} is a self-configuring ad hoc network architecture designed to provide a drop-in replacement for the grid's primary (\gls{ISP}) communication networks. The network requires minimal deployment configuration and offers essential services for human-to-human (voice, text) and device-to-device (\gls{SCADA}) coordination. Uniform hardware and software architecture allows rapid deployment from a storage facility to substations. Nodes are designed for compatibility with a variety of link technologies, e.g., radio, fiber, or powerline. \cref{fig:net-arch} illustrates the overall network architecture.

\gls{PhoenixSEN} consists of a designated \gls{CC} and Phoenix nodes deployed to substations or relay points. The nodes are interconnected with short-distance and long-distance links. Some of the links can be provided by third-parties, e.g., the National Guard. \gls{PhoenixSEN} is designed to be deployed into geographic areas served by multiple utility companies. The network provides an isolated virtual network to each utility on top of shared physical infrastructure. The virtual network spans all utility's substations. Each Phoenix node connect its substation \glspl{LAN} to the appropriate virtual network and also route packets for virtual networks of other utilities. The \gls{CC} provides additional equipment and services to facilitate network monitoring and management, and to support one-way broadcast communication across the deployment area.

Each Phoenix node provides multiple \glspl{VLAN} to its substation. Typically, there will be one \gls{VLAN} for \gls{SCADA} devices and another \gls{VLAN} for \gls{IT} (backend) systems. The addressing architecture of each \gls{VLAN} is configurable, allowing it to match the original \gls{ISP} network in order to minimize the need to reconfigure existing equipment. Each Phoenix node also provides essential network services locally to enable communication within the substation even when disconnected from \gls{PhoenixSEN}. These services include, among others, \gls{DNS}, \gls{DHCP}, \gls{NTP}, and \gls{VoIP} signaling.

The primary purpose of \gls{PhoenixSEN} is to restore connectivity in a grid under network-based cyber attack. We assume the grid's devices (\gls{SCADA} and \gls{IT}) might be compromised and may contribute to the attack. For this reason, Phoenix nodes provide a dedicated access port for a forensic team and services through which portions of the network or individual devices can be isolated from the rest of the network. \gls{PhoenixSEN} also comes with a built-in \gls{IDS} that attempts to automatically mitigate certain types of insider attacks.

\subsection{Phoenix Node}\label{sec:phoenix-node}

\begin{figure}
  \includegraphics[width=\linewidth]{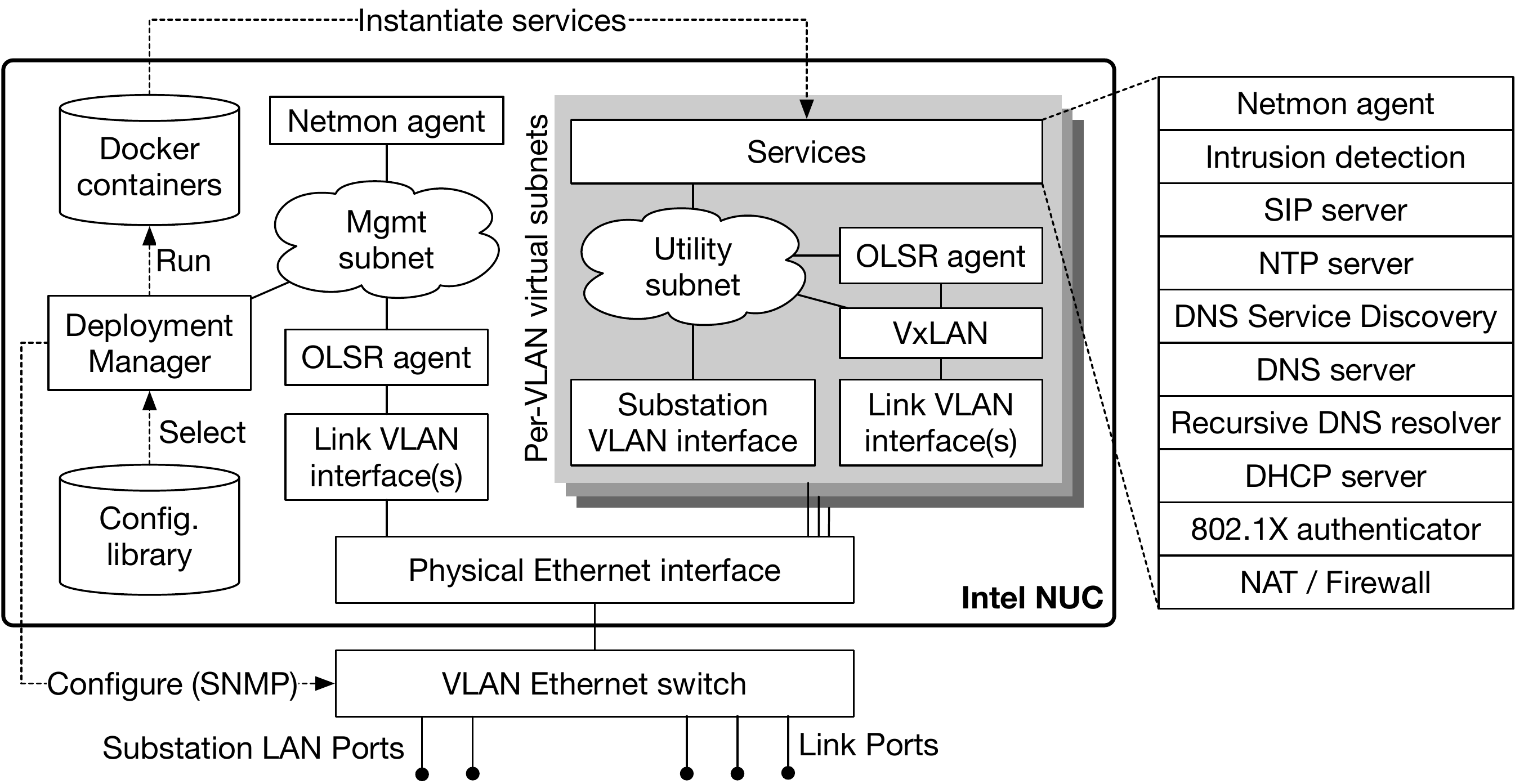}%
  \caption{Phoenix node software architecture. An isolated network environment (grey) with with all required services is created for each substation LAN. The environments that belong to the same utility are connected across PhoenixSEN.}\label{fig:phoenix-node-sw-arch}
\end{figure}

The Phoenix node is designed to be deployed from a storage facility to substations by ground transportation or via air lift shortly after a blackout. All hardware comes in a weather-resistant enclosure which contains all essential components such as Intel \gls{NUC}, Ethernet switches and cables, \gls{GPS}, Wi-Fi access points, and \gls{VoIP} clients. All Phoenix nodes use a uniform hardware and software architecture to simplify deployment and installation. \cref{fig:node-arch} illustrates the hardware architecture, \cref{fig:phoenix-node-sw-arch} provides an overview of the software architecture. \cref{fig:phoenix-sen-prototype} shows a small-scale prototype built for field evaluations.

The Phoenix node is based on the \href{https://www.intel.com/content/www/us/en/products/boards-kits/nuc.html}{Intel \gls{NUC}} small form-factor computer. The \gls{OS} is \href{https://ubuntu.com}{Ubuntu Linux 16.04} in a minimal configuration. All custom software for \gls{PhoenixSEN} is pre-installed in the form of \href{https://www.docker.com/}{Docker} containers. The containers are built from source code and are specifically designed to support re-building in the field without internet access, e.g., after modifications to fix critical bugs or vulnerabilities.

A newly deployed Phoenix node starts in a minimal configuration. In this state, the node starts only a deployment manager process and a management \gls{VLAN}. Utility and substation specific services are not yet provided. The deployment manager provides a \gls{UI} for the substation crew to perform one-time deployment configuration, i.e., to enter the utility name and substation number. The management \gls{VLAN} can be used by the \gls{CC} for remote support, configuration, or maintenance.

The configuration for utility and substation specific services is generated by a custom configuration synthesis program out of a model of the whole \gls{PhoenixSEN}. The model contains parameters such as number of utilities, number of substations per utility, desired \gls{IP} addressing architecture, types of \glspl{VLAN} and links, etc. In most cases the model can be created prior to \gls{PhoenixSEN} deployment based on information collected from utilities in the deployment region. In that case, the generated configuration library can be pre-loaded onto each Phoenix node. If the model is unavailable prior to deployment, the synthesis can be performed at the \gls{CC} and the generated configuration library can be distributed to substations on a \gls{USB} memory card, or the \gls{CC} can run the synthesis program on Phoenix nodes remotely over the management network.

Upon receiving the utility name and substation number, the node selects the appropriate configuration from the configuration library and starts utility and substation specific services and \glspl{VLAN}. A fully configured Phoenix node runs one or more isolated virtual network environments. The purpose of the network environment is to emulate the primary \gls{ISP} network that connected the substation to the internet before blackout. Nodes with multiple links also serve as transparent routers for other utilities sharing the \gls{PhoenixSEN}. The virtual network environments are configured for a particular substation and typically correspond to the substation's \glspl{VLAN}, e.g., \gls{IT} systems, \gls{VoIP} devices (phones), and \gls{SCADA}. Each environment is connected to a subset of the ports on the Ethernet switch. One port connects to the substation \gls{LAN}, the other ports connect to link modems or radios. The network environments are implemented using the Linux networking namespace capability with services provided by Docker containers.

Each network environment runs its own instance of the \gls{OLSR} agent which publishes the environment's \gls{IP} subnet information to \gls{PhoenixSEN}. This information is exchanged only with \gls{OLSR} agents that belong to the same utility and environment type (e.g., \gls{SCADA}). For example, utility A's \gls{SCADA} environment on one Phoenix node connects to utility A's \gls{SCADA} environments on all other Phoenix nodes, but not to utility B's \gls{SCADA} environments.

The utilities served by the same \gls{PhoenixSEN} can use conflicting or overlapping \gls{IP} subnets (e.g., 192.168.x.y). Supporting such scenarios natively would require \gls{VLAN} support on all \gls{PhoenixSEN} links. In order to stay compatible with a wide variety of link layer technologies (including \gls{IP}-only point-to-point links), the node does not require \gls{VLAN} support on links. Instead, it passes all traffic through a \gls{VxLAN}~\cite{rfc7348} gateway which encapsulates Ethernet frames in \gls{UDP} packets prior to transmission. A unique \gls{VxLAN} \gls{VNI} is generated for each utility-\gls{VLAN} combination during configuration synthesis.

Each network environment provides fully isolated elementary network services to the corresponding substation \gls{VLAN} such as \gls{DHCP}, \gls{DNS}, or \gls{NTP}. The \gls{DNS} service is described in \cref{sec:naming-discovery}. Each environment also runs a custom Netmon agent to discover and identify devices connected to the substation \gls{LAN}. The Netmon service is described in more detail in \cref{sec:netmon}.

The type of the network environment determines what additional network services are created. For example, the \gls{VoIP} environment provides a \gls{SIP}~\cite{rfc3261} server on each Phoenix node to keep the substation's phones operational and to provide a means of communication between the substation personnel and the \gls{CC}. Please refer to \cref{sec:voice-chat} for more detail. The \gls{SCADA} environment provides additional services for intrusion detection and mitigating insider attacks by compromised devices (\cref{sec:ethershield}).

\subsection{Deployment \& Network Formation}\label{sec:network-formation}

The deployment of \gls{PhoenixSEN} begins with the deployment of the \gls{CC}. The \gls{CC} then coordinates the deployment of the rest of the network and provides remote assistance to substations while they are setting up Phoenix nodes. During the initial deployment phase, the \gls{CC} may only have one-way broadcast capability, e.g., a high-power \gls{HF} radio with voice or low-speed data support. The \gls{CC} crew can use the broadcast channel to transmit substation-specific setup instructions, just enough to get the substation connected. Once the substation is connected, the \gls{CC} crew can help configure the substation's Phoenix node remotely.

Upon Phoenix node delivery, the substation crew begins setting up the node. We assume the crew has technical background (e.g., in electric grid engineering), but not necessarily in \gls{IT}, communications, or networking. The node includes a playbook and detailed installation instructions. All parts and connectors are clearly labeled. The instructions are detailed enough to allow the crew to independently setup the Phoenix node in a minimal configuration with a low-bandwidth control connection to the \gls{CC}. To facilitate this critical first step, the node enclosure may include a pre-configured \gls{HF} receiver to receive broadcast communication from the \gls{CC}.

When powered on, the Phoenix node begins to search for other Phoenix nodes on its link interfaces. Each interface has an instance of the \gls{OLSR} \href{http://www.olsr.org}{daemon} configured to automatically discover other \gls{OLSR}[-enabled]~\cite{rfc3626} nodes reachable over the interface. Gradually, Phoenix nodes form an \gls{OLSR}[-based] \gls{MANET} that eventually spans all substations and the \gls{CC}. Once the network is formed, the \gls{CC} can further configure and coordinate deployed Phoenix nodes over the network.

While forming, \gls{PhoenixSEN} provides connectivity in phases, gradually providing additional modes of communication as the system transitions from one phase to another. The phases are as follows.
\begin{enumerate}[leftmargin=0.2in]
  \item Low-speed, broadcast-only communication from the \gls{CC} to substations in the process of deploying a Phoenix node.
  \item Low-speed command and control connection between the \gls{CC} and each substation. The connection provides just enough bandwidth for the \gls{CC} crew to configure the Phoenix node remotely.
  \item The Phoenix node is fully connected to the network, but the overlay may not have yet fully formed or the node may be in the process of resolving an addressing conflict. The substation may not be able to reach all other substations of the same utility yet.
  \item The node is fully connected and provides \gls{VoIP} and text communication between the \gls{CC} and the substation crew.
\end{enumerate}

\noindent
Finally, the substation or \gls{CC} crew perform one-time deployment configuration by entering the utility name and substation number into the node. This step can be performed locally at the substation or remotely over the network. The  node uses the entered information to create utility and substation specific services and \glspl{VLAN}. The process is described in more detail in~\cref{sec:phoenix-node}.

\section{Naming and Service Discovery}\label{sec:naming-discovery}


For compatibility with existing devices and applications, \gls{PhoenixSEN} provides naming and service discovery services based on the \gls{DNS}. The requirements outlined in \cref{sec:motivation-problem-statement} and the fact that \gls{PhoenixSEN} must remain usable when disconnected from the internet make the traditional \gls{DNS} architecture with a single authoritative master server difficult to use in an ad hoc network. In \gls{PhoenixSEN}, each node should be able to independently contribute its own records into a flat namespace based on a \gls{DNS} zone shared by all nodes. In this section, we describe the design of a hybrid peer-to-peer \gls{DNS} system used in \gls{PhoenixSEN}.

The two primary use-cases for \gls{DNS} in \gls{PhoenixSEN} are discovering services offered by Phoenix nodes and supporting third-party tools with pre-existing \gls{TLS} certificates. The first use-case is implemented in the \gls{VoIP} and chat service, as described in \cref{sec:voice-chat}, which uses \gls{DNS} service discovery to map phone numbers to Phoenix nodes in a distributed manner. The second case refers to forensic and cyber-security activities performed with third-party tools. During live exercises, external participants often needed to be able to connect their tools with mandatory \gls{TLS} server certificate validation to \gls{PhoenixSEN}. The \gls{DNS} service allows registering domain names that match pre-existing \gls{TLS} certificates on a first-come first-serve basis and resolving those names across \gls{PhoenixSEN}.

The \gls{DNS} service takes advantage of the following: 1) a weak eventual consistency model is sufficient for \gls{DNS}; 2) in a network like \gls{PhoenixSEN}, the impact of conflicts and inconsistencies can be minimized network-wide through design. Point 1) expands the range of protocols that could be used for \gls{DNS} database replication to gossip (epidemic), flooding, and multicast-based protocols. Point 2) means that if applications and devices are configured appropriately, the probability of conflicts due to a weak consistency model will be negligible.

\gls{PhoenixSEN} provides such \gls{DNS} subsystem, with no dedicated master server and without any single point of failure. The subsystem is a hybrid of regular \gls{DNS} and multicast \gls{DNS}~\cite{rfc6762}. Devices and services can use the subsystem for network-wide \gls{DNS-SD}~\cite{rfc6763}. In an ad hoc network such as \gls{PhoenixSEN}, built-in service discovery provides a means for applications to discover services dynamically rather than relying on static configuration.

Dynamic service discovery improves robustness in scenarios where the network is impaired or only partially formed. A \gls{DNS}[-based] service discovery mechanism allows the reuse of higher layer protocols and services, e.g., \gls{TLS} server certificate validation.

\begin{figure}
  \includegraphics[width=\linewidth]{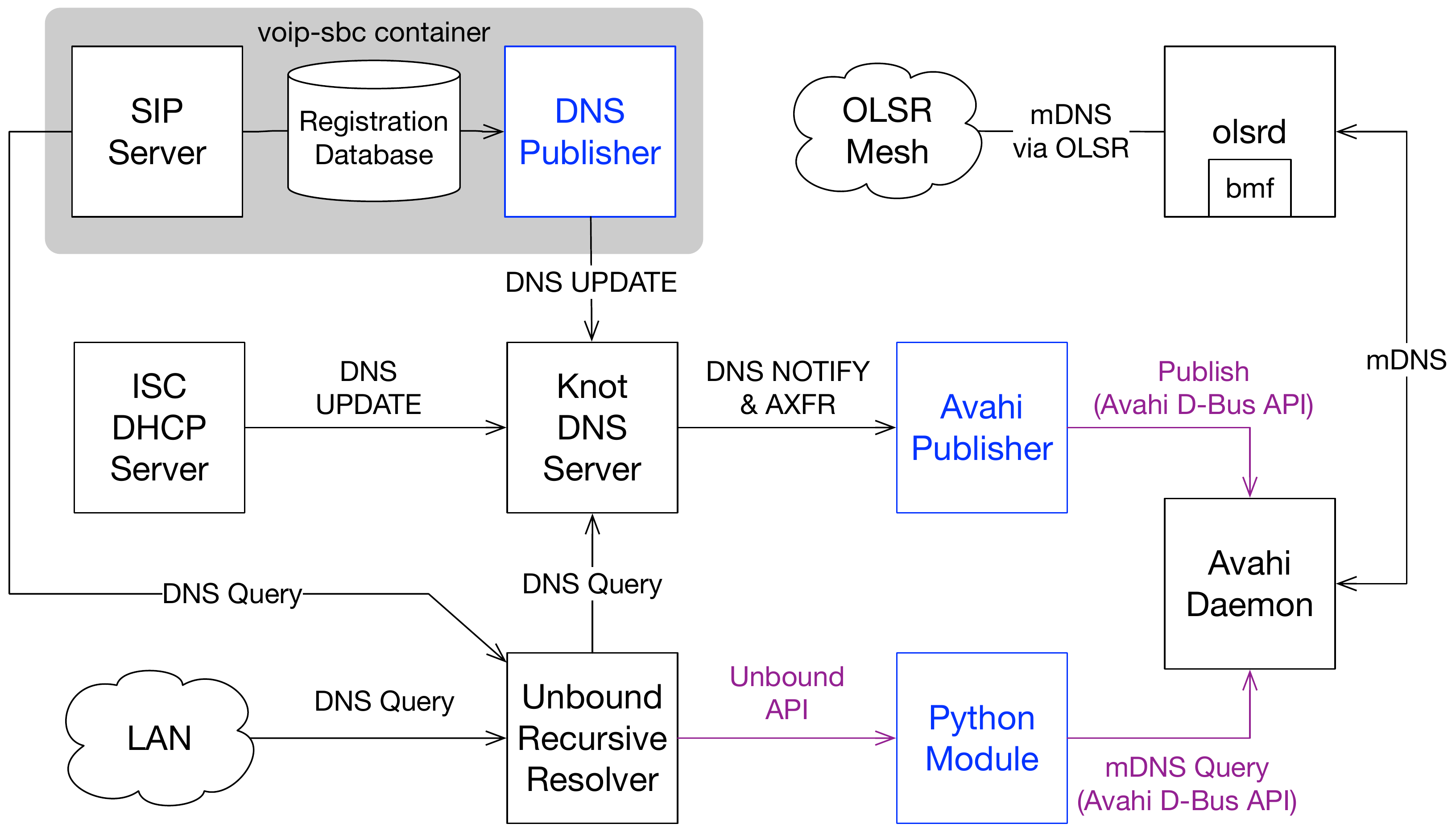}
  \caption{DNS subsystem architecture. Each Phoenix node runs the same set of services. The block with gray background shows the VoIP service using the DNS subsystem for service discovery. Custom-made blocks are highlighted in blue. Communication via proprietary protocols is highlighted in purple.}\label{fig:phoenix-dns}
\end{figure}

\cref{fig:phoenix-dns} illustrates the architecture of the \gls{DNS} subsystem. Each Phoenix node runs the full set of services that make up the \gls{DNS} subsystem. Thus, \gls{DNS} is always available on substation \glspl{LAN}, even if the Phoenix node itself is isolated from the rest of the network. As more Phoenix nodes join the network, \gls{DNS} records gathered from other nodes will become automatically available to the devices. Where possible, the subsystem uses standardized protocols and existing (unmodified) open source software.

Each Phoenix node provides a recursive \gls{DNS} resolver to substation \glspl{LAN} based on the \href{https://nlnetlabs.nl/projects/unbound}{Unbound resolver}. We chose Unbound because its behavior can be customized with an external Python program. We use that feature to resolve ordinary \gls{DNS} queries via \gls{mDNS}~\cite{rfc6762} (described later).

Unbound first forwards the query to a local authoritative \gls{DNS} server implemented with \href{https://www.knot-dns.cz}{Knot \gls{DNS} server}. The \gls{DNS} server is authoritative for the domain \texttt{phxnet.org} and for a portion of the \texttt{in-addr.arpa} space corresponding to the \gls{IP} subnet allocated to the Phoenix node. The domain \texttt{phxnet.org} represents a namespace shared by all Phoenix nodes. Any node can register a record in the shared namespace. If multiple nodes register the same record, all such records are merged when a client attempts to resolve the record. It is left up to the application to resolve the potential conflict.

If no record is found, Unbound forwards the query to an external \href{https://github.com/NLnetLabs/unbound/blob/master/pythonmod/examples/avahi-resolver.py}{Python module}. The module attempts to resolve the query via \gls{mDNS} using a local Avahi daemon instance\footnote{Avahi is the de facto standard implementation of \gls{mDNS} in Linux based OSes.}. The Python module and Avahi daemon communicate via a D-Bus \gls{API}.

In a typical configuration, the Avahi daemon multicasts \gls{DNS} queries over \gls{LAN} interfaces, e.g., local Ethernet or Wi-Fi interfaces. Since multicast \gls{DNS} packets are not forwarded by layer 3 routers, only nodes connected to the same link can be reached this way. \gls{PhoenixSEN} is, however, a routed network managed by an \gls{OLSR}~\cite{rfc3626} daemon running on each node. The daemon includes the \gls{BMF} plugin that enables multicast communication for the overlay network. We use the plugin to propagate Avahi's \gls{mDNS} packets across the \gls{OLSR} network. The \gls{BMF} plugin forwards multicast packets along a spanning tree covering the entire network, i.e., all Phoenix nodes.

The Avahi daemon on each Phoenix node publishes all \gls{DNS} records from the local \gls{DNS} server via \gls{mDNS}. This is accomplished with a custom program called ``Avahi Publisher''. The program behaves as a secondary \gls{DNS} server for the shared zone. When the zone is updated, Knot sends a DNS NOTIFY request to the program. The program issues a \gls{DNS} zone transfer to Knot to download all records and publishes those records via Avahi's D-BUS \gls{API}. Thus, \gls{LAN} clients that resolve \gls{DNS} records via Unbound receive records not only from the local Knot \gls{DNS} server, but also from all \gls{DNS} servers found anywhere in the network.

Devices that obtain an \gls{IP} address via \gls{DHCP} are automatically assigned hostnames by the \gls{DHCP} server. The \gls{DHCP} server uses the standard dynamic \gls{DNS} (DNS UPDATE) protocol to publish records to the local \gls{DNS} server. For example, a \gls{DHCP} client with the name ``foo'' will be assigned the hostname \texttt{foo.phxnet.org}. Having an automatically generated hostname for each device has two main benefits: 1) the device can be reached by its chosen name across the network; 2) services running on the device can use \gls{TLS} certificates with a wild-card certificate issued by \href{https://letsencrypt.org}{Letsencrypt}.

\begin{figure*}
  \includegraphics[width=\linewidth]{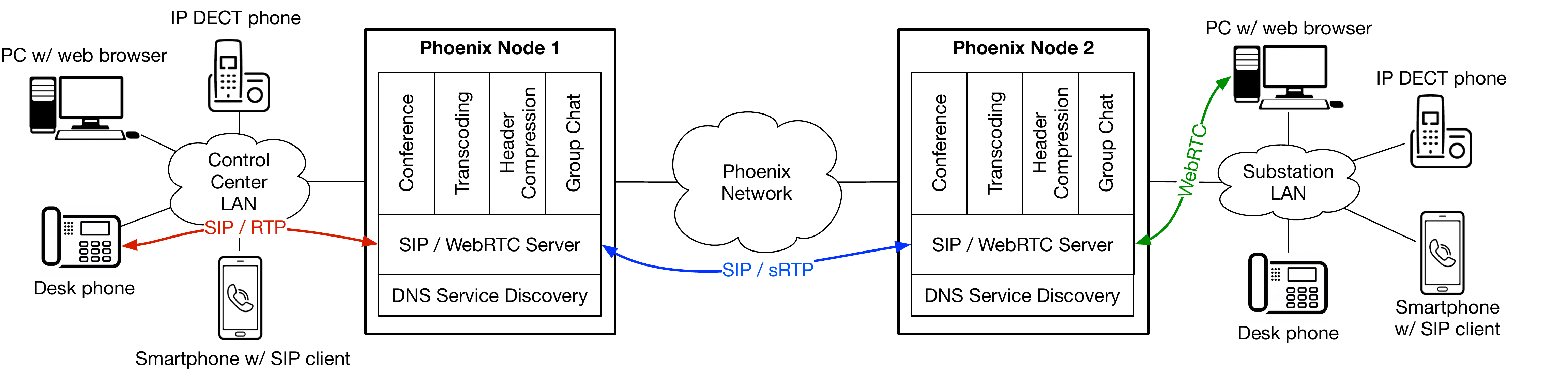}
  \caption{Internal architecture of the VoIP and chat subsystem. The subsystem can integrate a variety of SIP (red) and WebRTC (green) based VoIP clients. Each Phoenix node runs a full set of services. VoIP clients register at the nearest node. Communication sessions that traverse slow links (blue) are optionally transcoded and compressed. VoIP clients and servers rely on DNS-SD for server federation and client discovery. IP multicast (used by DNS-SD) helps make the subsystem fully decentralized and with no single point of failure.}\label{fig:voip-architecture}
\end{figure*}

\subsection{Service Discovery Example}\label{sec:discovery-example}

\gls{PhoenixSEN} provides a \gls{SIP}[-based] decentralized \gls{VoIP} service (highlighted in gray in \cref{fig:phoenix-dns}) designed to help substation personnel coordinate during a black-start recovery. The service relies on the \gls{PhoenixSEN} \gls{DNS} subsystem for network-wide service discovery. We discuss how the \gls{VoIP} service uses the \gls{DNS} subsystem in this section.

The \gls{VoIP} clients at a substation register with the nearest \gls{SIP} server. Since each Phoenix node runs a dedicated \gls{SIP} server, the \gls{VoIP} clients usually register with the \gls{SIP} at their substation's Phoenix node. To discover the \gls{SIP} server, a \gls{VoIP} client performs the standard procedure described in~\cite{rfc3263}. The \gls{DNS} queries issued by the \gls{VoIP} client will be forwarded by Unbound to the local Knot \gls{DNS} server. The \gls{DNS} server has the corresponding \gls{DNS} PTR and SRV resource records in its zone and those records point the \gls{VoIP} client to the \gls{SIP} server on the local Phoenix node. This way, all \gls{VoIP} clients across the whole network can have identical configuration and can, in case of smartphones, roam between substations.

Upon \gls{VoIP} client registration, the \gls{SIP} server publishes a custom \gls{DNS} record mapping the \gls{VoIP} client's number to the hostname of the \gls{SIP} server where it has registered, for example

{\small
\begin{verbatim}
4822._voip.phxnet.org. IN CNAME voip-phx23.phxnet.org.
\end{verbatim}}

\noindent
where 4822 is a VoIP client's phone number and \texttt{voip-phx23.phxnet.org} is the hostname of the Phoenix node where the client is registered. A custom program called ``DNS Publisher`` (\cref{fig:phoenix-dns}) publishes the record to the local \gls{DNS} server via \gls{DNS} UPDATE and the record is subsequently disseminated across the network with multicast \gls{DNS}.

When a remote \gls{SIP} server receives a \gls{SIP} INVITE~\cite{rfc3261} for the \gls{VoIP} client, it extracts the called number from the Request-URI and constructs a domain name from the number within the \texttt{\_voip.phxnet.org} suffix. The \gls{SIP} server then queries the \gls{DNS} for the corresponding CNAME record. If a match is found, the call is forwarded to the \gls{SIP} identified by the record.

Note that the multicast \gls{DNS} subsystem is configured to pro-actively disseminate the requests across the network. Thus, that for existing (registered) numbers, the query will usually be answered by the local \gls{DNS} server from its cache fast. Queries for non-existing (unregistered) numbers take a few seconds until multicast \gls{DNS} times out.

Both \gls{VoIP} clients and servers rely on the \gls{DNS} service discovery feature to locate each other. This helps make the \gls{VoIP} architecture fully decentralized and scalable, allows (mobile) \gls{VoIP} clients to roam between Phoenix nodes, and makes swapping hardware components, e.g., \gls{VoIP} clients, easy in the case of compromise or malfunction.

\section{Voice and Chat}\label{sec:voice-chat}

One of the purposes of \gls{PhoenixSEN} is to facilitate the coordination of people during an emergency. To that end, the network provides built-in support for \gls{SIP}[-based]~\cite{rfc3261} real-time voice and text communication. The service supports the following modalities: two-party calls, multi-party conferencing, text messaging, and multi-party group chat (with a persistent log). \cref{fig:voip-architecture} shows the internal architecture of the subsystem.

The \gls{VoIP} subsystem has been designed to support a variety of \gls{VoIP} clients. \gls{IP} \gls{DECT} phone represents \gls{VoIP} clients that come bundled with the Phoenix node hardware. Each Phoenix node comes with a couple of \gls{IP} \gls{DECT} phones pre-configured for use with the network. The client labeled as ``PC with a web browser'' could be any tablet, laptop, or a PC with a recent web-browser with \gls{WebRTC} support. The \gls{SIP} server includes a JavaScript \gls{WebRTC} application that can turn any such device into a fully-featured \gls{VoIP} and chat client without the need to install any software. The desk phone icon represents \gls{VoIP} clients that had existed at the substation before the Phoenix node was installed. The smartphone \gls{VoIP} client represents mobile Android devices with a \gls{SIP} client application installed. The mobile clients are connected via a Wi-Fi network created by the Phoenix node and can roam between substations while keeping the same number.

The \gls{VoIP} and chat subsystem uses a simple four-number dialing plan where each substation is allocated a fixed prefix. The substation's \gls{VoIP} clients are assigned numbers from that prefix. Each Phoenix node runs a full set of \gls{VoIP} services (\gls{SIP} and \gls{WebRTC} servers, conference server, chat server). The clients register at the nearest \gls{SIP} server, typically the one located on the Phoenix node installed in the client's substation. Having a full set of \gls{VoIP} services on each Phoenix node allows the node to function in isolation. Even if the node is disconnected from \gls{PhoenixSEN}, calls and chats within its substation should remain possible.

The subsystem supports ordinary two-way calls between any two \gls{VoIP} clients, multi-party conference calls with conference rooms created on demand, peer-to-peer chat on compatible devices (Android clients), and a web-based group chat service accessible through the JavaScript \gls{WebRTC} client. The \gls{VoIP} service supports persistent logging and archiving of conversations (both voice and text), if needed. The \gls{SIP} server on each Phoenix node performs transparent media transcoding and enforces authentication and encryption on calls that traverse \gls{PhoenixSEN}, i.e., two or more Phoenix nodes. Transcoding helps establish an upper bound on the \gls{VoIP} bandwidth on slow (long-haul) links and improves robustness of calls traversing such links.

A simple four-digit dialing plan with prefixes allocated to substations was used during live exercises, however, the \gls{VoIP} subsystem has no fixed dialing plan hard-coded. Instead, it relies on \gls{DNS-SD} to discover available \gls{SIP} servers and \gls{VoIP} clients, and to discover the \gls{SIP} server where a particular \gls{VoIP} client is registered. \cref{sec:discovery-example} describes in detail how the \gls{VoIP} subsystem uses \gls{DNS-SD}. Having a fully dynamic \gls{VoIP} subsystem has important benefits: 1) Phoenix nodes have uniform hardware and software configuration and are interchangeable; 2) mobile \gls{VoIP} clients roaming from one substation to another are supported; and 3) the subsystem is fully decentralized with no single point of failure.


\section{Network Monitoring}\label{sec:netmon}

\begin{figure}
  \includegraphics[width=\linewidth]{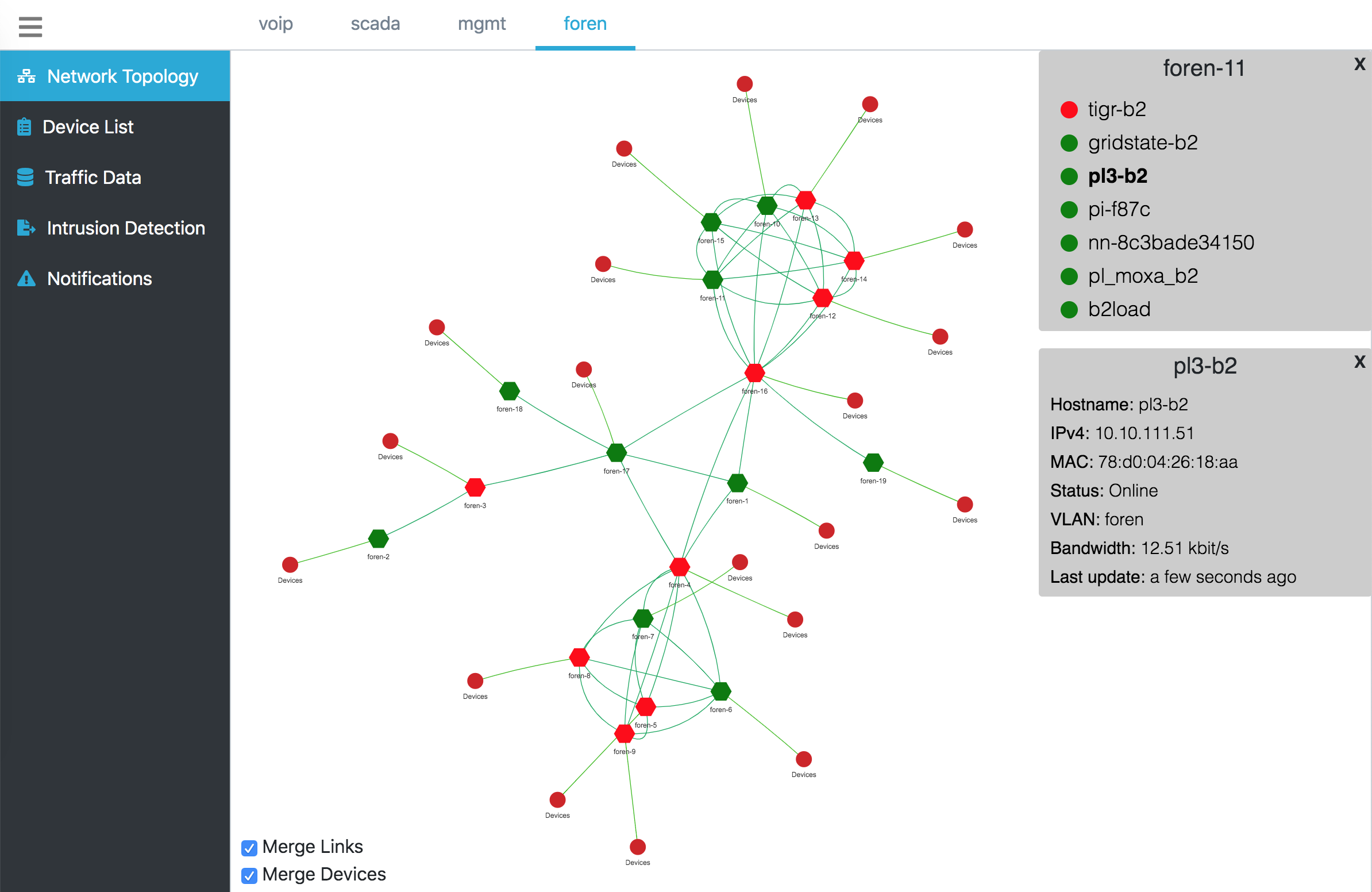}
  \caption{A PhoenixSEN topology graph shown by Netmon during a live exercise. Elements that might require attention are highlighted in red. Hexagonal nodes represent substation Phoenix nodes. The menus on the right show a list of a substation's devices (top) and detail about a particular device (bottom).}\label{fig:netmon-topology}
\end{figure}

In a geographically distributed ad hoc network architecture such as \gls{PhoenixSEN}, real-time network monitoring and situation awareness is key for successful deployment and operation. In this section, we describe Netmon, a network monitoring service designed and developed for \gls{PhoenixSEN}. Netmon provides near real-time information about the state of the network through a web-based \gls{UI}. The \gls{UI} lets the control center see the topology of the formed \gls{OLSR} network and provides alerts when it detects security-related events and incidents. \cref{fig:netmon-topology} shows the \gls{UI} displaying the network topology graph of a \gls{PhoenixSEN} prototype during a live exercise.

\begin{figure}
  \includegraphics[width=\linewidth]{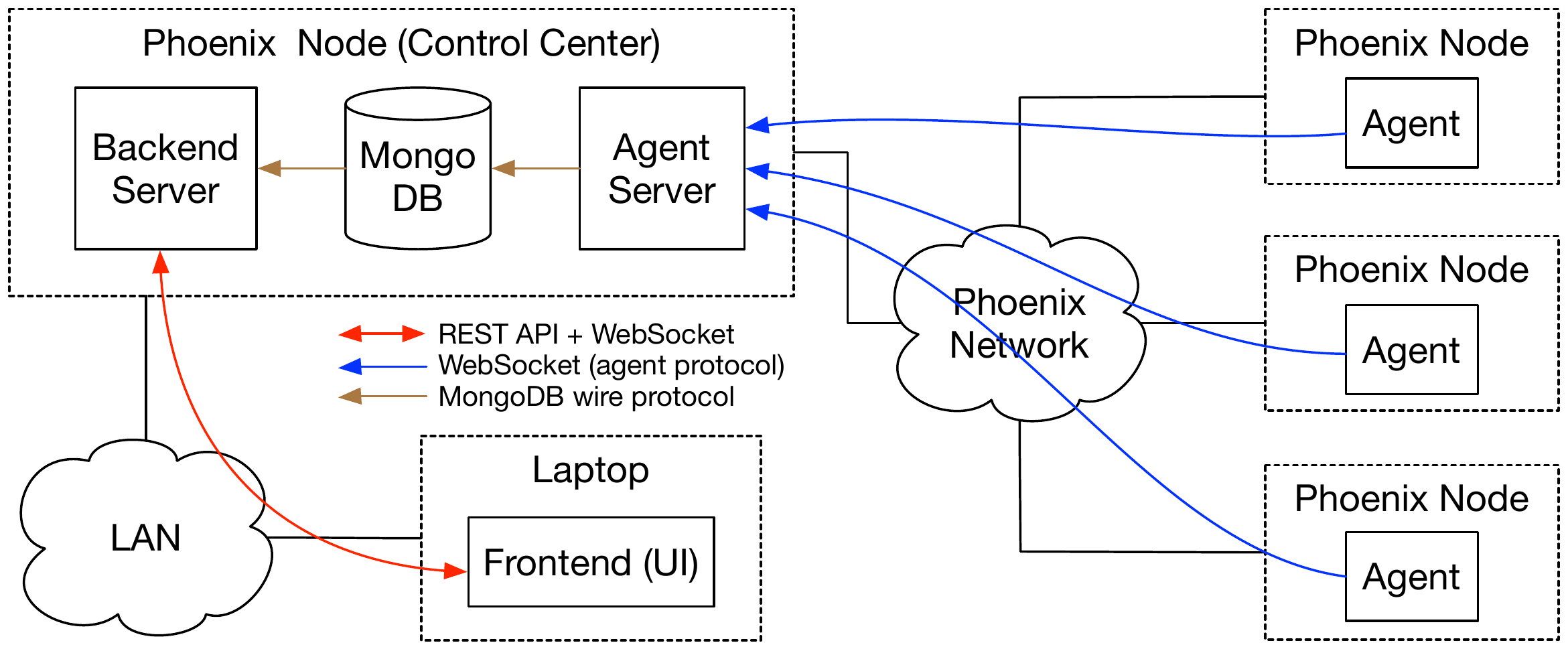}
  \caption{Netmon internal architecture. Each Phoenix node runs an agent process that streams collected data to a backend server in the control center. Network operators interact with Netmon via a web-based UI implemented in JavaScript. The architecture provides a near real-time overview of the network.}\label{fig:netmon-architecture}
\end{figure}

The internal architecture of Netmon is shown in \cref{fig:netmon-architecture}. Every Phoenix node runs a custom Netmon agent process. The agent collects data about the state of the node using \gls{OS}[-level] instrumentation and about directly attached substation (\gls{LAN}) devices using active network scanning. It also provides an \gls{API} for other processes on the same node, e.g., an \gls{IDS}. The collected data includes statistics from selected network interfaces, the state of the node's \gls{OLSR} links, and information about discovered devices.

To discover devices attached to the node's substation \gls{LAN} interfaces, the agent periodically issues \gls{ARP} requests for all \gls{IP} addresses that belong to the interface's configured \gls{IP} subnet. A device that responds to the \gls{ARP} request is recorded and will be displayed in the network topology graph, as shown in \cref{fig:netmon-topology}. Once a \gls{LAN} device has been discovered, it is periodically contacted to determine the device's reachability. The agent also probes the state of selected \gls{UDP} and \gls{TCP} ports on all discovered \gls{LAN} devices to see what services might be offered by the device.

A separate \gls{IDS} process, running on the same Phoenix node, provides additional data to the Netmon agent. The data represents potential intrusion detection incidents and security threats. Netmon transmits the data to the backend where it is then presented to the network operator by the frontend (\gls{UI}).

Each agent maintains a persistent WebSocket~\cite{rfc6455} connection to an agent server discovered via \gls{DNS} SRV. The control center Phoenix node provides one agent server instance for each \gls{VLAN}. As long as the agent is connected to the server, data is streamed to the agent server in near real-time. When an agent gets disconnected from the server, it temporarily stores the collected data in a local cache. All locally cached data will be uploaded to the server later once the agent has reconnected.

The agent server stores all the data from all connected agents in a persistent database. The data is processed and indexed to allow time-based addressing and aggregation, i.e., retrieving the state of the network at a particular time. This feature allows debugging or post-mortem analysis after an exercise when the physical network infrastructure is no longer operational. A backend server, also running on the control center Phoenix node, serves the data to the frontend (\gls{UI}) via a RESTful and WebSocket based \gls{API}. The \gls{API} allows the frontend code to retrieve any data generated by the agents, and to receive asynchronous (push) updates as new data becomes available.

The \gls{UI} is a JavaScript application running in a web browser on a laptop in the control center. The application is specifically designed to always show the most recent network state without the need to reload the browser window. The \gls{UI} is automatically updated based on push notifications received from the backend server. The \gls{UI} provides a quick ``at a glance'' overview of the entire network, as well as detailed information about individual network components. The operator can use the UI to quickly scan the network for faulty (red) nodes and links, or see which parts of the network require immediate attention.

The Netmon agent is implemented in Python, uses \href{https://scapy.net}{Scapy} for network probing and device discovery, and \href{https://www.sqlite.org}{SQLite} for the local cache. The agent and backend servers are both implemented in JavaScript running in \href{https://nodejs.org}{NodeJS}. All collected data is stored in a \href{https://mongodb.com}{MongoDB} database on the control center Phoenix node. The frontend is a JavaScript application implemented with \href{https://vuejs.org}{Vue.js} and running in a web browser.

\section{Mitigating Insider Attacks}\label{sec:ethershield}

Insider attacks are of particular concern in large physical infrastructures such as the electric grid. An attacker could use compromised devices in a substation \gls{LAN} to attempt to thwart recovery attempts. If the attacker also gains physical access to the substation, they could plant a malicious device in the substation's network (``device-in-a-closet'')~\cite{rpi-in-closet} and use the device to launch \gls{MITM} attacks. Lack of Ethernet security combined with ever shrinking form factor of embedded devices makes such attacks practical and very hard to discover.

\begin{figure}
  \includegraphics[width=\linewidth]{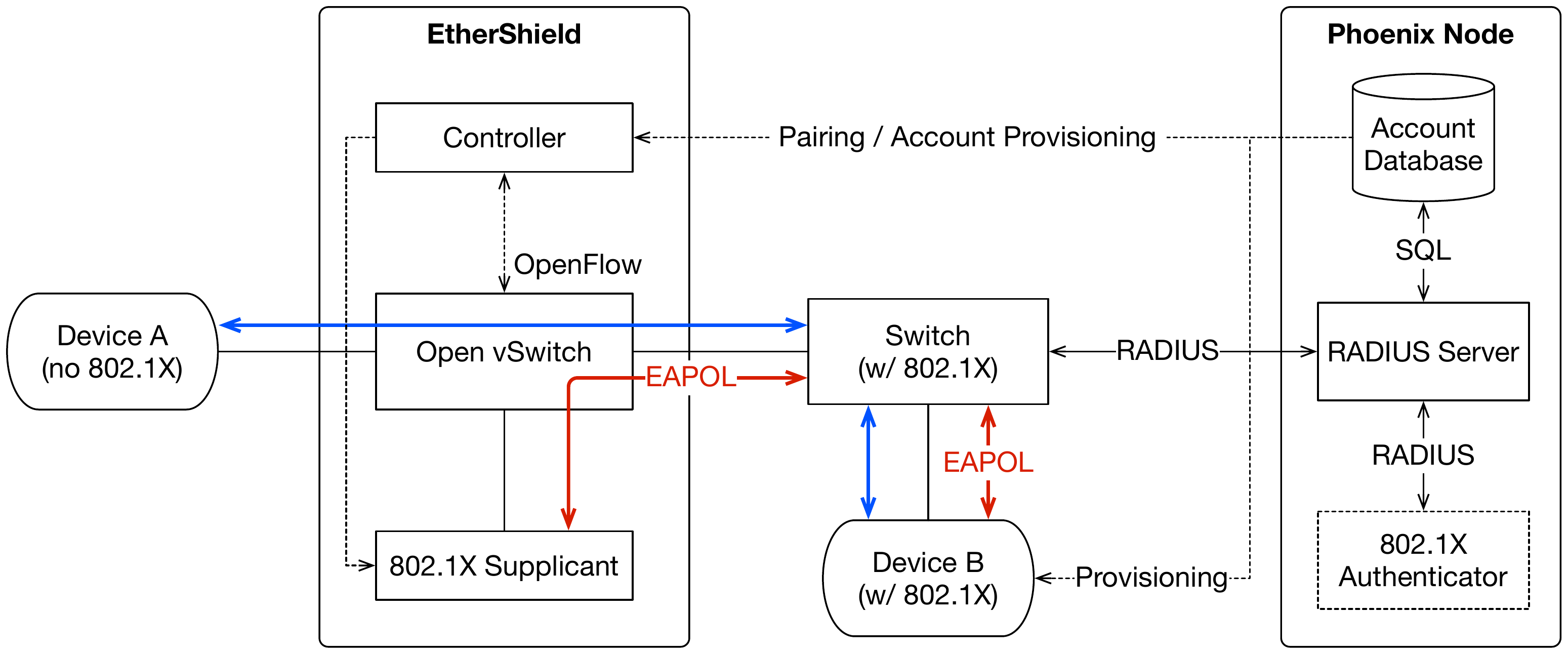}
  \caption{The architecture of the EtherShield subsystem. An Ethernet switch-like device is inserted between a trusted SCADA device and untrusted network. The device transparently authenticates all Ethernet frames received and sent by the device. The diagram shows one possible configuration for networks that support Port-based Network Access Control (IEEE 802.1X).}\label{fig:ethershield}
\end{figure}

To help mitigate the risk of network-based insider attacks, we designed and built a prototype device called EtherShield. EtherShield is a ``bump in the wire'' embedded device with two Ethernet ports: internal and external. The internal port is connected to a trusted (not compromised) \gls{SCADA} device, preferably, as close to the \gls{SCADA} device as possible to minimize the risk of malicious actors getting access to that segment of the network. The external port is connected to the untrusted substation \gls{LAN}. Until activated, the device operates as a regular Ethernet switch. Once activated from the nearest Phoenix node, the device transparently authenticates all communication between the \gls{SCADA} device and the network. \cref{fig:ethershield} shows the architecture of the EtherShield subsystem.

Each EtherShield device will have been paired with the nearest Phoenix node via a \gls{MITM}[-resistant] \gls{USB} connection prior to deployment into the \gls{LAN}. During pairing, the EtherShield device and the Phoenix node generate the cryptographic material for secure communication over an untrusted \gls{LAN}. Until activated, the EtherShield device passes all traffic between its two ports unmodified. Once a \gls{SCADA} device protected with an EtherShield is deemed secure, the network operator can activate a secure mode in the EtherShield to isolate the \gls{SCADA} device from potentially malicious traffic. In a secure mode, the EtherShield transparently authenticates all packets between the \gls{SCADA} device, the nearest Phoenix node, and (optionally) other devices within the substation \gls{LAN}.

The authentication is based on either IEEE 802.1X~\cite{8021x} or IPsec~\cite{rfc6701}. The EtherShield can be configured to only let authenticated traffic through to the \gls{SCADA} device, or to also allow traffic from other \gls{LAN} devices not protected by an EtherShield device. This gives the network operator some flexibility while incrementally securing the substation \gls{SCADA} network. Critical \gls{SCADA} devices can be incrementally protected and isolated from malicious traffic without disrupting the operation of the rest of the substation \gls{LAN}.

We built a prototype EtherShield device based on the \href{https://www.raspberrypi.org}{Raspberry Pi} running \href{https://www.openvswitch.org/}{Open vSwitch}. In the default ``open'' mode of operation, Open vSwitch is configured to function as a simple learning Ethernet switch. When switched into a secure mode, the device starts either an IEEE 802.1X supplicant or an IPsec tunnel and re-configures Open vSwitch to redirect a portion of the traffic to those services. For example, in the IEEE 802.1X mode only \gls{EAPoL} frames are redirected to the supplicant process. Frames received from the protected device are transparently augmented with an authentication header (IEEE 802.1X or IPsec authentication header) as they pass through the switch. A frame received from the untrusted \gls{LAN} for the protected device is first authenticated before the frame is forwarded to the device. A custom controller application provides a secure \gls{HTTP} \gls{API}. This \gls{API} can be used to configure and control the device from the nearest Phoenix node.

The Phoenix node provides a standards-based set of services to support the EtherShield subsystem. The cryptographic material generated during pairing is kept in a \href{https://www.postgresql.org}{PostgreSQL} database, along with account information. The node provides an IEEE 802.1X authenticator to the substation \gls{LAN} based on the \gls{RADIUS} protocol~\cite{rfc2865}. The authentication is implemented with \href{https://freeradius.org}{FreeRADIUS}. An IPsec-based authenticator based on \href{https://strongswan.org}{strongSwan} is also provided.

\section{Related Work}\label{sec:related-work}

\textbf{Disaster Communications}. The critical role of communication in the aftermath of a large-scale disaster event became apparent after hurricane Katrina in 2005~\cite{failure-of-initiative}, hurricane Irene in 2011, and during the 2017 Atlantic hurricane season~\cite{connectingdots}. In Puerto Rico, over 11 million people lost electricity for 11 days in the largest blackout in \gls{US} history~\cite{vox}. These events rendered existing communication networks across large geographic areas unusable~\cite{katrina-lessons-learned,fcc-hurricane-impact}. Additionally, significant challenges caused by non-interoperable communication systems were reported~\cite{nist-disaster}, limiting situational awareness and preventing communication across jurisdictions and organizations~\cite{stute2020empirical,comfort-katrina}. A 2016 \gls{NIST} report~\cite{Appl16:Critical} found communication failures to be a major obstacle in all recent disaster recovery efforts.

\textbf{Industrial Control Systems}. Modern industrial control systems, including the power grid, require communication to function. As critical infrastructure, such systems are increasingly targets of cyber attacks~\cite{obama-remarks,nescor}. A 2011 vulnerability analysis performed by the Idaho National Laboratory found that applying traditional \gls{IT} system protective measures to real-time energy delivery control systems is inadequate and could lead to a power disruption~\cite{idaho}. The Stuxnet worm~\cite{stuxnet} and the Maroochy water breach~\cite{slay2007lessons} incidents illustrate the danger network-based cyber attacks pose for industrial control systems. Even ordinary planned upgrades can have devastating consequences in systems that control physical processes~\cite{merrimack}.

\textbf{Network Architectures}. A network architecture suitable for emergency scenarios will inevitably be ad hoc and temporary~\cite{redcross}. Such architectures have been the subject of active research for decades and many promising ideas have been proposed including peer-to-peer, mesh, mobile, delay-tolerant, opportunistic, and hybrid network architectures~\cite{30years,kumar2010current,gotenna}. Despite the wide range of communication architectures, basic interoperability between organizations is still a problem.

There have been attempts to repurpose existing technology to provide services where communication networks are unavailable. The Village Telco project~\cite{village-telco} designed and built a \gls{DIY} toolkit for voice and text communications based on wireless mesh networking and \gls{VoIP}. The Osmocom project~\cite{osmocom} provides the essential building blocks for temporary cellular network infrastructures.

\textbf{Disaster Management Systems}. Recently, specialized disaster management systems have been gaining interest. Several organizations specializing in emergency response have made their solutions available~\cite{redcross,nethope,ares,sahana}. The core functionality of these systems centers around crowd-sourced field data gathering, incident tracking, and visualization. Such systems typically place no special requirements on the underlying communication networks. There have been attempts to use advanced technology in emergencies, for example, Panacea Glass proposes to use Google Glass and cloud computing for situation awareness and effective triage of patients~\cite{gillis2015panacea,panacea-cloud}.

\textbf{Network Monitoring Systems}. The ad hoc nature of temporary emergency networks calls for some form of real-time monitoring for topology discovery, bandwidth estimation, intrusion detection, and troubleshooting in the field. A vast number of network measurement and monitoring tools have been developed~\cite{nms-list}. The tools can be broadly classified into passive, active, and hybrid (a combination of active and passive)~\cite{passive-active-monitoring}.

Passive monitoring relies on packet capture~\cite{mccanne1993bsd,tcpdump}, \gls{OS} instrumentation~\cite{rfc3549}, or sampling~\cite{rfc3176} to perform measurements. Passive methods require existing network traffic to function, the ability to observe data flows within the network, and are generally less intrusive than active methods~\cite{john2010passive}. Notable passive monitoring tools include CoralReef~\cite{coralreef}, Bro~\cite{paxon1998using}, Wireshark~\cite{wireshark}, Snort~\cite{snort}, and sFlow~\cite{sflow}.

Active monitoring estimates network properties by observing the handling of special-purpose data injected into the network by the monitoring tool. Active methods are generally intrusive, generate variable amounts of artificial network load, and may interfere with application traffic. Existing active methods rely on \gls{IP} options (\gls{ICMP} echo~\cite{rfc792}, traceroute~\cite{rfc1393}), \gls{TCP} congestion control (iPerf~\cite{iperf}), bandwidth and transmission time probing (BWPing~\cite{bwping}), or packet dispersion measurements~\cite{dovrolis2001packet,capprobe}. Scapy~\cite{scapy} is a popular packet manipulation tool for active network discovery and monitoring.

An important aspect of any network monitoring system is its data processing architecture. The architecture determines the number and locations of collection points, types of collected data, collection protocols, and data processing algorithms. If data collection takes place over the network being monitored, care must be taken to minimize the generated overhead and its impact on the monitored network. Well-known monitoring data collection protocols include Cisco NetFlow~\cite{rfc3954}, the IPFix protocol~\cite{rfc7011}, and \gls{SNMP}~\cite{rfc1157}. \gls{SNMP} is widely supported by existing networking equipment.

The following monitoring tools inspired Netmon, the monitoring tool presented in this paper: Moloch~\cite{moloch}, MRTG~\cite{mrtg}, OpenNMS~\cite{opennms}, Cacti~\cite{cacti}, and Zabbix~\cite{zabbix}.

\section{Conclusion}\label{sec:conclusion}

We presented the design and prototype implementation of \gls{PhoenixSEN}, an ad hoc network architecture specifically designed to enable real-time coordination of people and (\gls{SCADA}) devices during an emergency, while the primary network infrastructure may be inoperable. Our network architecture is designed with sufficient flexibility to be used a drop-in replacement for network infrastructure and services typically provided by third-party \glspl{ISP}. Our work is primarily motivated by the needs of the power distribution industry during a hypothetical large-scale blackout triggered by a network-based cyber attack. We believe \gls{PhoenixSEN} has the potential to speed up power grid recovery and service restoration, in particular during a cyber attack that is persistent and ongoing.

\begin{figure*}
  \subfloat{\includegraphics[width=0.21\linewidth]{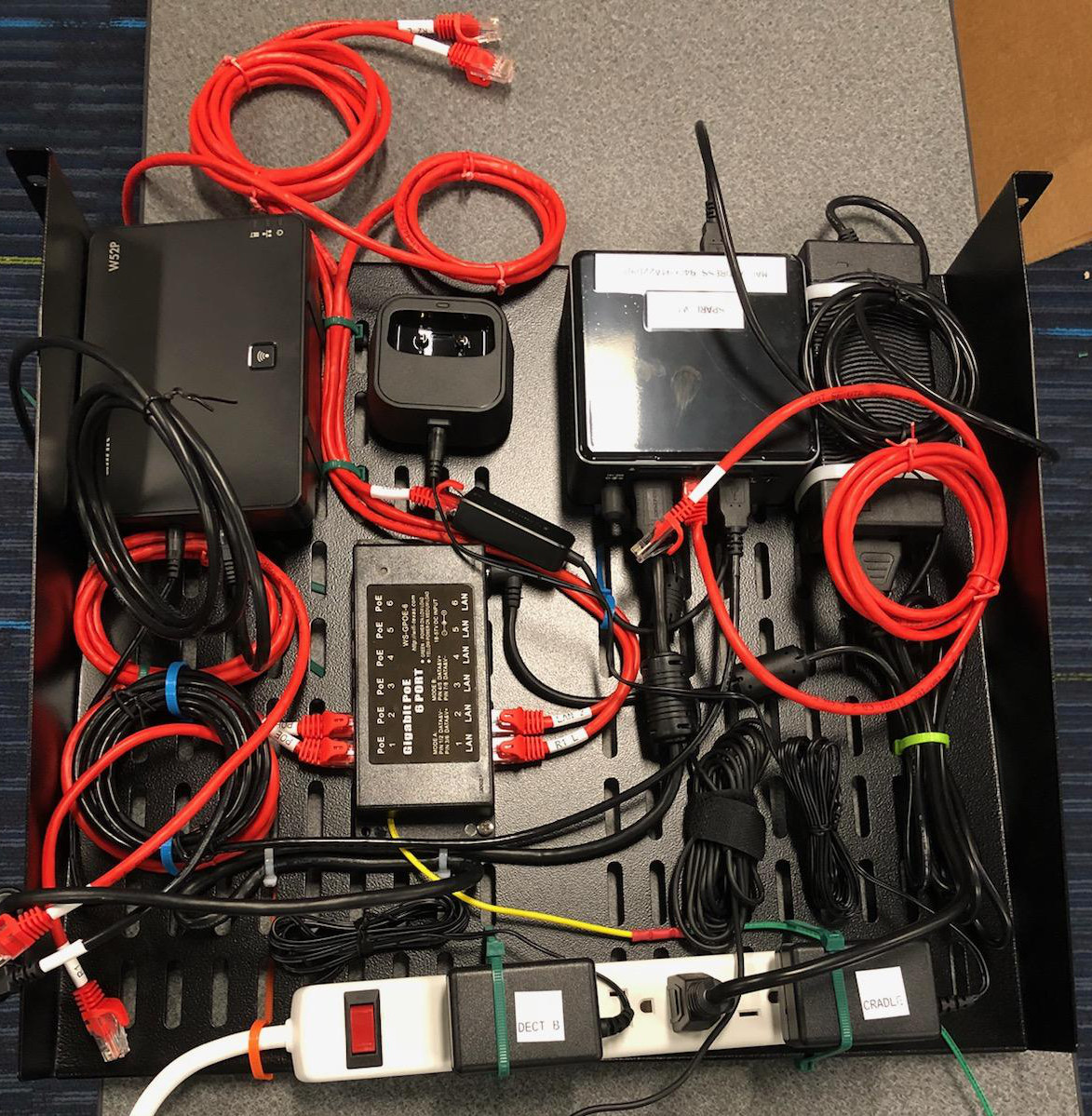}}
  \hspace*{\fill}
  \subfloat{\includegraphics[width=0.35\linewidth]{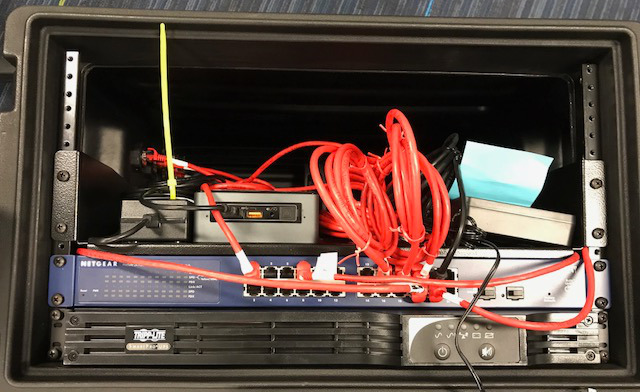}}
  \hspace*{\fill}
  \subfloat{\includegraphics[width=0.43\linewidth]{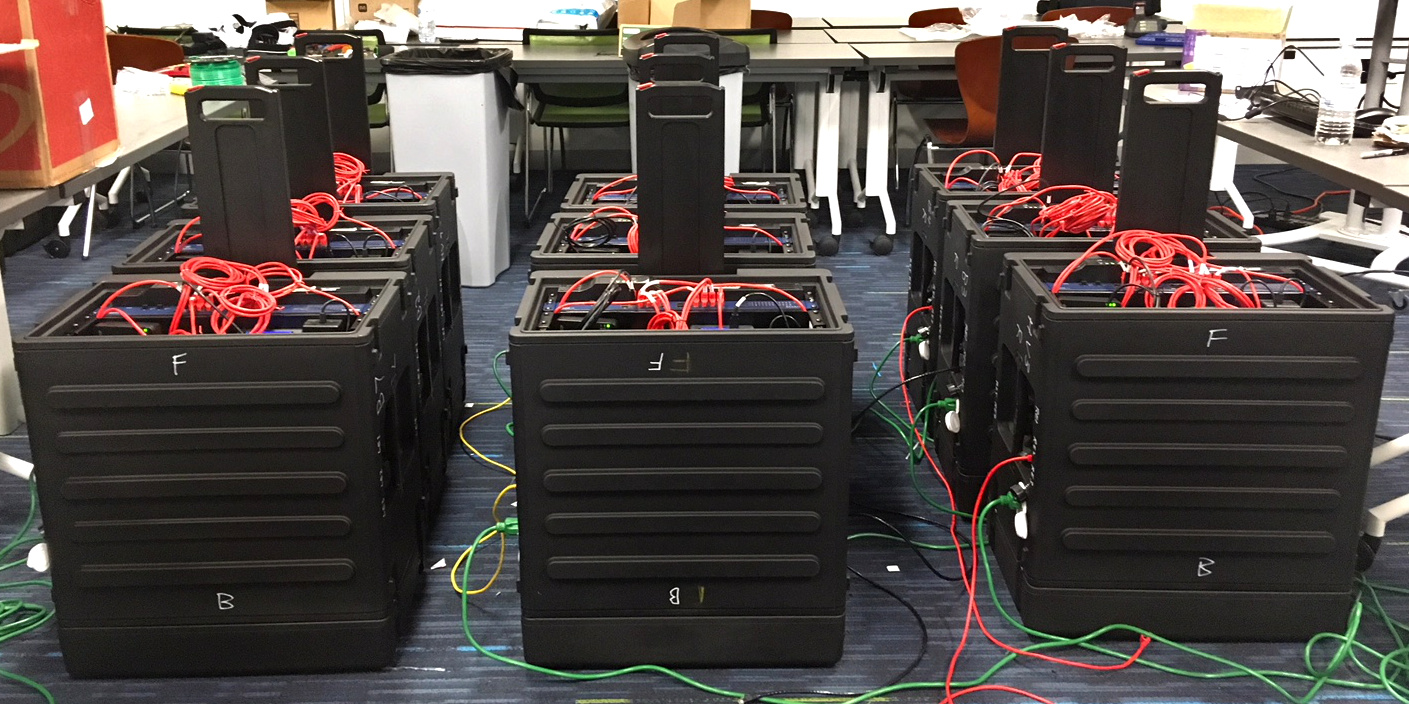}}
  \caption{Photos of the PhoenixSEN prototype evaluated through DARPA-led exercises. Left to right: Phoenix node components (Intel NUC, POE switch, Yealink W52P phone), Phoenix node in weather-resistant enclosure (Intel NUC, Yealink W52P phone, Netgear Ethernet switch, Tripp Lite UPS), small-scale Phoenix SEN prototype installation in a lab. The photos were taken by Hema Retty at the University of Illinois at Urbana-Champaign in September 2018.}\label{fig:phoenix-sen-prototype}
\end{figure*}

Several iterations of the \gls{PhoenixSEN} prototype described in this paper were tested and evaluated through a series of \gls{DARPA}[-led] cyber security exercises on Plum Island, NY~\cite{plum-island}. During the exercise, \gls{PhoenixSEN} was used together with a variety of power grid recovery tools contributed by other exercise participant to asses the readiness of the infrastructure to recover from a simulated large-scale cyber attack on the electrical grid. Photos in \cref{fig:phoenix-sen-prototype} show the evaluated \gls{PhoenixSEN} prototype.

This paper describes the most recent iteration of \gls{PhoenixSEN}. Some of the features described in this paper ended up not being evaluated in live exercises. The EtherShield prototype device described in \cref{sec:ethershield} was developed and tested in a lab at Columbia University. The hardware deployed on Plum Island did not have \gls{GPS} receivers since the precise location of each Phoenix node was known in advance. One-time deployment configuration (\cref{sec:phoenix-node}) was performed from a laptop attached to the Phoenix node via an Ethernet port. The configuration synthesis approach was only tested on a Columbia University testbed. The chat feature (\cref{sec:voice-chat}), which was originally developed to support team coordination during the exercise, was eventually replaced with a third-party application the participants were already familiar with.


\subsection{Future Work}\label{sec:future-work}

Although the design of \gls{PhoenixSEN} was primarily motivated by specific needs of the power distribution industry, we believe the resulting architecture is more general and flexible, and might be suitable for other emergency scenarios as well. As part of future work, \gls{PhoenixSEN} could be extended with features and services for first responders, disaster recovery, and other emergency scenarios where communication and coordination is critical even if the primary communication networks are not operational. We envision supporting a variety of common emergency and disaster scenarios on top a general-purpose hardware and software architecture with interchangeable service profiles. Each profile will activate services designed to meet the needs of a particular emergency scenario, e.g., incident management, ticketing, collaborative document editing, and others.

The \gls{VoIP} subsystem in \gls{PhoenixSEN} provides all the usual modes of real-time communication: two-way calls, multi-party conference calls, direct chat, and group chat. It is likely that \gls{PhoenixSEN} might get deployed in challenging scenarios where ordinary calls and chats may not be the most efficient means of communication. As part of an effort to support more emergency scenarios, further research into alternative means of (near) real-time communication might be warranted. Specifically, good support for \gls{PTT} communication seems to be a promising direction for first responder communication systems.

Recent advances in automated speech-to-text and text-to-speech technologies might provide a means to automatically transcribe and index emergency voice communications. In addition, such technologies might provide better support for voice communication over long-haul and slow links such as \gls{HF} band links, satellite links, and \gls{PLC} systems. A combination of \gls{RoHC}~\cite{rfc5795}, \gls{PTT}, transcoding, and voice-to-text could be potentially used to design a near real-time, highly robust, and resilient communication system suitable for very challenged communication networks and scenarios.

We plan to continue developing the Netmon software with features specifically designed for mobile ad hoc networks. We believe that no other existing network monitoring solution supports ad hoc or highly dynamic networks well. In a future version of Netmon, we plan to remove the restriction to have only one instance of backend services per network. We will aim to make Netmon fully decentralized, with the backend services and data sharded across all Phoenix nodes. To build a global view of the network, Netmon will obtain and merge data obtained from all Phoenix nodes.

\section*{Acknowledgment}\label{sec:acknowledgment}
\addcontentsline{toc}{section}{Acknowledgment}

The work described in this paper was very much a team effort. The authors developed the \gls{PhoenixSEN} prototype as part of a joint team with members from BAE Systems and Perspecta Labs.



We would also like to thank Clifton Lin, Charles Tao, Defu Li, Ranga Reddy, and James Dolan, all members of the BAE Systems team, for fruitful discussions and help in developing and testing the prototype.

We are indebted to Frafos GmbH, the developers of the \href{https://www.frafos.com/abc-sbc}{ABC \gls{SBC}}, who generously provided us with a free license to use the \gls{SBC} in our prototype. Furthermore, Frafos engineers were instrumental in helping us to install and configure the \gls{SBC} during the early stages of our project.



%

\bibliographystyle{IEEEtran}
\bibliography{bibs/references,bibs/rfc}

\begin{thebibliography}{10}
\providecommand{\url}[1]{#1}
\csname url@samestyle\endcsname
\providecommand{\newblock}{\relax}
\providecommand{\bibinfo}[2]{#2}
\providecommand{\BIBentrySTDinterwordspacing}{\spaceskip=0pt\relax}
\providecommand{\BIBentryALTinterwordstretchfactor}{4}
\providecommand{\BIBentryALTinterwordspacing}{\spaceskip=\fontdimen2\font plus
\BIBentryALTinterwordstretchfactor\fontdimen3\font minus
  \fontdimen4\font\relax}
\providecommand{\BIBforeignlanguage}[2]{{%
\expandafter\ifx\csname l@#1\endcsname\relax
\typeout{** WARNING: IEEEtran.bst: No hyphenation pattern has been}%
\typeout{** loaded for the language `#1'. Using the pattern for}%
\typeout{** the default language instead.}%
\else
\language=\csname l@#1\endcsname
\fi
#2}}
\providecommand{\BIBdecl}{\relax}
\BIBdecl

\bibitem{argentina2019}
\BIBentryALTinterwordspacing
{CNN}. (2019, Aug.) {Massive failure leaves Argentina, Paraguay and Uruguay
  with no power}. [Online]. Available:
  \url{https://edition.cnn.com/2019/06/16/world/power-outage-argentina-uruguay-paraguay}
\BIBentrySTDinterwordspacing

\bibitem{uk2019}
\BIBentryALTinterwordspacing
{BBC}. (2019, Aug.) {Major power failure affects homes and transport}.
  [Online]. Available: \url{https://www.bbc.com/news/uk-49300025}
\BIBentrySTDinterwordspacing

\bibitem{NationalGridESO}
\BIBentryALTinterwordspacing
{National Grid ESO}. Black start. [Online]. Available:
  \url{https://www.nationalgrideso.com/balancing-services/system-security-services/black-start}
\BIBentrySTDinterwordspacing

\bibitem{PJM12}
\BIBentryALTinterwordspacing
PJM. {PJM} manual 12: Balancing operations. [Online]. Available:
  \url{https://www.pjm.com/~/media/documents/manuals/m12.ashx}
\BIBentrySTDinterwordspacing

\bibitem{nist-smartgrid-roadmap}
\BIBentryALTinterwordspacing
{National Institute of Standards and Technology}, ``{NIST Framework and Roadmap
  for Smart Grid Interoperability Standards, Release 3.0},'' Sep. 2014.
  [Online]. Available: \url{http://dx.doi.org/10.6028/NIST.SP.1108r3}
\BIBentrySTDinterwordspacing

\bibitem{radics}
\BIBentryALTinterwordspacing
{Defense Advanced Research Projects Agency (DARPA)}. {Rapid Attack Detection,
  Isolation, and Characterization Systems (RADICS) program}. [Online].
  Available:
  \url{https://www.darpa.mil/program/rapid-attack-detection-isolation-and-characterization-systems}
\BIBentrySTDinterwordspacing

\bibitem{idaho}
{Idaho National Laboratory}. (2011, Sep.) {Vulnerability Analysis of Energy
  Delivery Control Systems (INL/EXT-10-18381)}.

\bibitem{uk2019b}
\BIBentryALTinterwordspacing
{National Grid ESO}. (2019, Aug.) {Interim Report into the Low Frequency Demand
  Disconnection (LFDD) following Generator Trips and Frequency Excursion on 9
  Aug 2019}. [Online]. Available:
  \url{https://www.nationalgrideso.com/document/151081/download}
\BIBentrySTDinterwordspacing

\bibitem{nist-disaster}
\BIBentryALTinterwordspacing
{Applied Technology Council}, ``{Critical Assessment of Lifeline System
  Performance: Understanding Societal Needs in Disaster Recovery},'' {National
  Institute of Standards and Technology}, Tech. Rep., Apr. 2016. [Online].
  Available: \url{http://dx.doi.org/10.6028/NIST.GCR.16-917-39}
\BIBentrySTDinterwordspacing

\bibitem{fcc-hurricane-impact}
\BIBentryALTinterwordspacing
{FCC}. (2018, Aug.) {2017 Atlantic Hurricane Season Impact on Communications
  Report and Recommendations Public Safety Docket No. 17-344}. [Online].
  Available: \url{https://docs.fcc.gov/public/attachments/DOC-353805A1.pdf}
\BIBentrySTDinterwordspacing

\bibitem{redcross}
\BIBentryALTinterwordspacing
{Red Cross}. (2019, Aug.) {Red Cross Disaster Services Technology}. [Online].
  Available: \url{http://www.tucsoncert.info/Red_Cross_DST_slides.pdf}
\BIBentrySTDinterwordspacing

\bibitem{ferc-889}
\BIBentryALTinterwordspacing
F.~E.~R. Commission, ``{Order No. 889: Open Access Same-Time Information System
  and Standards of Conduct},'' 1996. [Online]. Available:
  \url{https://www.ferc.gov/legal/maj-ord-reg/land-docs/order889.asp}
\BIBentrySTDinterwordspacing

\bibitem{DNP3}
``{IEEE Standard for Electric Power Systems Communications-Distributed Network
  Protocol (DNP3)},'' \emph{{IEEE Std 1815-2012 (Revision of IEEE Std
  1815-2010)}}, pp. 1--821, 2012.

\bibitem{ICCN}
{International Electrotechnical Commission}, ``{Telecontrol equipment and
  systems - Part 6-503: Telecontrol protocols compatible with ISO standards and
  ITU-T recommendations - TASE.2 Services and protocol},'' \emph{{IEC Std
  60870-6-503 TASE.2 Services and protocol}}, 2014.

\bibitem{iec-61850}
------, ``{Communication networks and systems for power utility automation},''
  \emph{{IEC Std 61850}}, 2013.

\bibitem{ieee-c37}
``{IEEE Standard for Synchrophasor Measurements for Power Systems},''
  \emph{{IEEE Std C37.118.1-2011}}, 2011.

\bibitem{rfc7348}
\BIBentryALTinterwordspacing
M.~Mahalingam, D.~Dutt, K.~Duda, P.~Agarwal, L.~Kreeger, T.~Sridhar,
  M.~Bursell, and C.~Wright, ``{Virtual eXtensible Local Area Network (VXLAN):
  A Framework for Overlaying Virtualized Layer 2 Networks over Layer 3
  Networks},'' RFC 7348, Internet Engineering Task Force, Aug. 2014. [Online].
  Available: \url{http://www.ietf.org/rfc/rfc7348.txt}
\BIBentrySTDinterwordspacing

\bibitem{rfc3261}
\BIBentryALTinterwordspacing
J.~Rosenberg, H.~Schulzrinne, G.~Camarillo, A.~Johnston, J.~Peterson,
  R.~Sparks, M.~Handley, and E.~Schooler, ``{SIP: Session Initiation
  Protocol},'' RFC 3261, Internet Engineering Task Force, Jun. 2002. [Online].
  Available: \url{http://www.ietf.org/rfc/rfc3261.txt}
\BIBentrySTDinterwordspacing

\bibitem{rfc3626}
\BIBentryALTinterwordspacing
T.~Clausen and P.~Jacquet, ``{Optimized Link State Routing Protocol (OLSR)},''
  RFC 3626, Internet Engineering Task Force, Oct. 2003. [Online]. Available:
  \url{http://www.ietf.org/rfc/rfc3626.txt}
\BIBentrySTDinterwordspacing

\bibitem{rfc6762}
\BIBentryALTinterwordspacing
S.~Cheshire and M.~Krochmal, ``{Multicast DNS},'' RFC 6762, Internet
  Engineering Task Force, Feb. 2013. [Online]. Available:
  \url{http://www.ietf.org/rfc/rfc6762.txt}
\BIBentrySTDinterwordspacing

\bibitem{rfc6763}
\BIBentryALTinterwordspacing
------, ``{DNS-Based Service Discovery},'' RFC 6763, Internet Engineering Task
  Force, Feb. 2013. [Online]. Available:
  \url{http://www.ietf.org/rfc/rfc6763.txt}
\BIBentrySTDinterwordspacing

\bibitem{rfc3263}
\BIBentryALTinterwordspacing
J.~Rosenberg and H.~Schulzrinne, ``{Session Initiation Protocol (SIP): Locating
  SIP Servers},'' RFC 3263, Internet Engineering Task Force, Jun. 2002.
  [Online]. Available: \url{http://www.ietf.org/rfc/rfc3263.txt}
\BIBentrySTDinterwordspacing

\bibitem{rfc6455}
\BIBentryALTinterwordspacing
I.~Fette and A.~Melnikov, ``{The WebSocket Protocol},'' RFC 6455, Internet
  Engineering Task Force, Dec. 2011. [Online]. Available:
  \url{http://www.ietf.org/rfc/rfc6455.txt}
\BIBentrySTDinterwordspacing

\bibitem{rpi-in-closet}
\BIBentryALTinterwordspacing
C.~Haschek. The curious case of the raspberry pi in the network closet.
  [Online]. Available:
  \url{https://blog.haschek.at/2018/the-curious-case-of-the-RasPi-in-our-network.html}
\BIBentrySTDinterwordspacing

\bibitem{8021x}
\BIBentryALTinterwordspacing
{IEEE}. (2020) {802.1X: Port-Based Network Access Control}. [Online].
  Available: \url{https://1.ieee802.org/security/802-1x/}
\BIBentrySTDinterwordspacing

\bibitem{rfc6701}
\BIBentryALTinterwordspacing
A.~Farrel and P.~Resnick, ``{Sanctions Available for Application to Violators
  of IETF IPR Policy},'' RFC 6701, Internet Engineering Task Force, Aug. 2012.
  [Online]. Available: \url{http://www.ietf.org/rfc/rfc6701.txt}
\BIBentrySTDinterwordspacing

\bibitem{rfc2865}
\BIBentryALTinterwordspacing
C.~Rigney, S.~Willens, A.~Rubens, and W.~Simpson, ``{Remote Authentication Dial
  In User Service (RADIUS)},'' RFC 2865, Internet Engineering Task Force, Jun.
  2000. [Online]. Available: \url{http://www.ietf.org/rfc/rfc2865.txt}
\BIBentrySTDinterwordspacing

\bibitem{failure-of-initiative}
\BIBentryALTinterwordspacing
{U.S. House of Representatives}. (2006) {A Failure of Initiative: Final Report
  of the Select Bipartisan Committee to Investigate the Preparation for and
  Response to Hurricane Katrina}. [Online]. Available:
  \url{https://www.congress.gov/congressional-report/109th-congress/house-report/377/1}
\BIBentrySTDinterwordspacing

\bibitem{connectingdots}
\BIBentryALTinterwordspacing
{Fitzpatrick, Leo and Scurato, Carmen and Torres, Joseph}. (2019, May)
  {Connecting the Dots. The Telecommunications Crisis in Puerto Rico}. {Free
  Press}. [Online]. Available:
  \url{https://www.freepress.net/sites/default/files/2019-05/connecting_the_dots_the_telecom_crisis_in_puerto_rico_free_press.pdf}
\BIBentrySTDinterwordspacing

\bibitem{vox}
\BIBentryALTinterwordspacing
{Campbel, Alexia Fernandez}. {It took 11 months to restore power to Puerto
  Rico}. {Vox}. [Online]. Available:
  \url{https://www.vox.com/identities/2018/8/15/17692414/puerto-rico-power-electricity-restored-hurricane-maria}
\BIBentrySTDinterwordspacing

\bibitem{katrina-lessons-learned}
\BIBentryALTinterwordspacing
{Federal Government of the United States}. (2006) {The Federal Response to
  Hurricane Katrina: Lessons Learned}. [Online]. Available:
  \url{https://georgewbush-whitehouse.archives.gov/reports/katrina-lessons-learned/}
\BIBentrySTDinterwordspacing

\bibitem{stute2020empirical}
M.~Stute, M.~Maass, T.~Schons, M.-A. Kaufhold, C.~Reuter, and M.~Hollick,
  ``Empirical insights for designing information and communication technology
  for international disaster response,'' \emph{International Journal of
  Disaster Risk Reduction}, p. 101598, 2020.

\bibitem{comfort-katrina}
\BIBentryALTinterwordspacing
L.~K. Comfort and T.~W. Haase, ``{Communication, Coherence, and Collective
  Action: The Impact of Hurricane Katrina on Communications Infrastructure},''
  \emph{Public Works Management \& Policy}, vol.~10, no.~4, pp. 328--343, 2006.
  [Online]. Available: \url{https://doi.org/10.1177/1087724X06289052}
\BIBentrySTDinterwordspacing

\bibitem{Appl16:Critical}
\BIBentryALTinterwordspacing
L.~Johnson, T.~D. O'Rourke, S.~Chang, C.~A. Davis, L.~D.-O.~I. Robertson,
  H.~Schulzrinne, and K.~Tierney, ``Critical assessment of lifeline system
  performance: Understanding societal needs in disaster recovery,'' National
  Institute of Standards and Technology, Gaithersburg, MD, Report NIST CGR
  16-917-39, Apr. 2016. [Online]. Available:
  \url{https://www.nist.gov/sites/default/files/documents/el/resilience/NIST-GCR-16-917-39.pdf}
\BIBentrySTDinterwordspacing

\bibitem{obama-remarks}
\BIBentryALTinterwordspacing
{The White House}. (2009) {Remarks by the President on Securing Our Nation's
  Cyber Infrastructure}. [Online]. Available:
  \url{https://obamawhitehouse.archives.gov/the-press-office/remarks-president-securing-our-nations-cyber-infrastructure}
\BIBentrySTDinterwordspacing

\bibitem{nescor}
\BIBentryALTinterwordspacing
{National Electric Sector Cybersecurity Organization Resource (NESCOR}.
  {Analysis of Selected Electric Sector High Risk Failure Scenarios}. [Online].
  Available: \url{https://smartgrid.epri.com/NESCOR.aspx}
\BIBentrySTDinterwordspacing

\bibitem{stuxnet}
S.~{Karnouskos}, ``Stuxnet worm impact on industrial cyber-physical system
  security,'' in \emph{IECON 2011 - 37th Annual Conference of the IEEE
  Industrial Electronics Society}, Nov 2011, pp. 4490--4494.

\bibitem{slay2007lessons}
J.~Slay and M.~Miller, ``{Lessons Learned from the Maroochy Water Breach},'' in
  \emph{International Conference on Critical Infrastructure Protection}.\hskip
  1em plus 0.5em minus 0.4em\relax Springer, 2007, pp. 73--82.

\bibitem{merrimack}
\BIBentryALTinterwordspacing
{National Transportation Safety Board}. (2019, Aug.) {Over-pressure of a
  Columbia Gas of Massachusetts Low-pressure Natural Gas Distribution System}.
  [Online]. Available:
  \url{https://www.ntsb.gov/investigations/AccidentReports/Pages/PLD18MR003-preliminary-report.aspx}
\BIBentrySTDinterwordspacing

\bibitem{30years}
F.~Legendre, T.~Hossmann, F.~Sutton, and B.~Plattner, ``30 years of wireless ad
  hoc networking research: what about humanitarian and disaster relief
  solutions? what are we still missing?'' in \emph{Proceedings of the 1st
  International Conference on Wireless Technologies for Humanitarian Relief
  (ACWR’11)}, 2011, pp. 217--217.

\bibitem{kumar2010current}
G.~V. Kumar, Y.~V. Reddyr, and D.~M. Nagendra, ``{Current research work on
  routing protocols for MANET: a literature survey},'' \emph{international
  Journal on computer Science and Engineering}, vol.~2, no.~03, pp. 706--713,
  2010.

\bibitem{gotenna}
\BIBentryALTinterwordspacing
{goTenna}. (2019, Aug.) {goTenna Mesh Network}. [Online]. Available:
  \url{https://gotenna.com/}
\BIBentrySTDinterwordspacing

\bibitem{village-telco}
M.~Adeyeye and P.~Gardner-Stephen, ``{The Village Telco project: a reliable and
  practical wireless mesh telephony infrastructure},'' \emph{EURASIP Journal on
  Wireless Communications and Networking}, vol. 2011, no.~1, p.~78, 2011.

\bibitem{osmocom}
\BIBentryALTinterwordspacing
{The Osmocom Project}. (2020) {Open Source Mobile Communications}. [Online].
  Available: \url{http://osmocom.org/}
\BIBentrySTDinterwordspacing

\bibitem{nethope}
\BIBentryALTinterwordspacing
{NetHope, Inc.} {NetHope}. [Online]. Available: \url{https://nethope.org}
\BIBentrySTDinterwordspacing

\bibitem{ares}
\BIBentryALTinterwordspacing
{American Radio Relay League}. {Amateur Radio Emergency Service (ARES)}.
  [Online]. Available: \url{http://www.arrl.org/ares}
\BIBentrySTDinterwordspacing

\bibitem{sahana}
\BIBentryALTinterwordspacing
{Sahana Foundation}. (2020) {Open Source Disaster Management Software}.
  [Online]. Available: \url{https://sahanafoundation.org/}
\BIBentrySTDinterwordspacing

\bibitem{gillis2015panacea}
J.~Gillis, P.~Calyam, A.~Bartels, M.~Popescu, S.~Barnes, J.~Doty, D.~Higbee,
  and S.~Ahmad, ``{Panacea's glass: Mobile cloud framework for communication in
  mass casualty disaster triage},'' in \emph{2015 3rd IEEE International
  Conference on Mobile Cloud Computing, Services, and Engineering}.\hskip 1em
  plus 0.5em minus 0.4em\relax IEEE, 2015, pp. 128--134.

\bibitem{panacea-cloud}
D.~{Jiang}, R.~{Huang}, P.~{Calyam}, J.~{Gillis}, O.~{Apperson},
  D.~{Chemodanov}, F.~{Demir}, and S.~{Ahmad}, ``Hierarchical cloud-fog
  platform for communication in disaster incident coordination,'' in \emph{2019
  7th IEEE International Conference on Mobile Cloud Computing, Services, and
  Engineering (MobileCloud)}, April 2019, pp. 1--7.

\bibitem{nms-list}
\BIBentryALTinterwordspacing
{Les Cottrell}. (2020) {Network Monitoring Tools}. [Online]. Available:
  \url{https://www.slac.stanford.edu/xorg/nmtf/nmtf-tools.html}
\BIBentrySTDinterwordspacing

\bibitem{passive-active-monitoring}
\BIBentryALTinterwordspacing
B.~B. Lowekamp, ``Combining active and passive network measurements to build
  scalable monitoring systems on the grid,'' \emph{SIGMETRICS Perform. Eval.
  Rev.}, vol.~30, no.~4, p. 19–26, Mar. 2003. [Online]. Available:
  \url{https://doi.org/10.1145/773056.773061}
\BIBentrySTDinterwordspacing

\bibitem{mccanne1993bsd}
S.~McCanne and V.~Jacobson, ``The bsd packet filter: A new architecture for
  user-level packet capture,'' in \emph{USENIX winter}, vol.~46, 1993.

\bibitem{tcpdump}
\BIBentryALTinterwordspacing
{The Tcpdump Group}. (2020) {tcpdump and libpcap}. [Online]. Available:
  \url{https://www.tcpdump.org/}
\BIBentrySTDinterwordspacing

\bibitem{rfc3549}
\BIBentryALTinterwordspacing
J.~Salim, H.~Khosravi, A.~Kleen, and A.~Kuznetsov, ``{Linux Netlink as an IP
  Services Protocol},'' RFC 3549, Internet Engineering Task Force, Jul. 2003.
  [Online]. Available: \url{http://www.ietf.org/rfc/rfc3549.txt}
\BIBentrySTDinterwordspacing

\bibitem{rfc3176}
\BIBentryALTinterwordspacing
P.~Phaal, S.~Panchen, and N.~McKee, ``{InMon Corporation's sFlow: A Method for
  Monitoring Traffic in Switched and Routed Networks},'' RFC 3176, Internet
  Engineering Task Force, Sep. 2001. [Online]. Available:
  \url{http://www.ietf.org/rfc/rfc3176.txt}
\BIBentrySTDinterwordspacing

\bibitem{john2010passive}
W.~John, S.~Tafvelin, and T.~Olovsson, ``Passive internet measurement: Overview
  and guidelines based on experiences,'' \emph{Computer Communications},
  vol.~33, no.~5, pp. 533--550, 2010.

\bibitem{coralreef}
K.~Keys, D.~Moore, R.~Koga, E.~Lagache, M.~Tesch, and k.~claffy, ``{The
  architecture of CoralReef: an Internet traffic monitoring software suite},''
  in \emph{Passive and Active Network Measurement Workshop (PAM)}.\hskip 1em
  plus 0.5em minus 0.4em\relax Amsterdam, Netherlands: RIPE NCC, Apr 2001.

\bibitem{paxon1998using}
V.~Paxon, ``Using bro to detect network intruders: experiences and status,'' in
  \emph{Proceedings of First International Workshop on the Recent Advances in
  Intrusion Detection}, 1998.

\bibitem{wireshark}
\BIBentryALTinterwordspacing
{The Wireshark Foundation}. (2020) {Wireshark}. [Online]. Available:
  \url{https://www.wireshark.org/}
\BIBentrySTDinterwordspacing

\bibitem{snort}
\BIBentryALTinterwordspacing
{The Snort Team}. (2020) {Snort: Network Intrusion Detection and Prevention
  System}. [Online]. Available: \url{https://www.snort.org/}
\BIBentrySTDinterwordspacing

\bibitem{sflow}
\BIBentryALTinterwordspacing
{InMon Corp.} (2020) {sFlow}. [Online]. Available: \url{https://sflow.org/}
\BIBentrySTDinterwordspacing

\bibitem{rfc792}
\BIBentryALTinterwordspacing
J.~Postel, ``{Internet Control Message Protocol},'' RFC 792, Internet
  Engineering Task Force, Sep. 1981. [Online]. Available:
  \url{http://www.ietf.org/rfc/rfc792.txt}
\BIBentrySTDinterwordspacing

\bibitem{rfc1393}
\BIBentryALTinterwordspacing
G.~Malkin, ``{Traceroute Using an IP Option},'' RFC 1393, Internet Engineering
  Task Force, Jan. 1993. [Online]. Available:
  \url{http://www.ietf.org/rfc/rfc1393.txt}
\BIBentrySTDinterwordspacing

\bibitem{iperf}
\BIBentryALTinterwordspacing
{iPerf Authors}. (2020) {iPerf}. [Online]. Available: \url{https://iperf.fr/}
\BIBentrySTDinterwordspacing

\bibitem{bwping}
\BIBentryALTinterwordspacing
{BWPing Authors}. (2020) {BWPing}. [Online]. Available:
  \url{https://bwping.sourceforge.io/}
\BIBentrySTDinterwordspacing

\bibitem{dovrolis2001packet}
C.~Dovrolis, P.~Ramanathan, and D.~Moore, ``What do packet dispersion
  techniques measure?'' in \emph{Proceedings IEEE INFOCOM 2001. Conference on
  Computer Communications. Twentieth Annual Joint Conference of the IEEE
  Computer and Communications Society (Cat. No. 01CH37213)}, vol.~2.\hskip 1em
  plus 0.5em minus 0.4em\relax IEEE, 2001, pp. 905--914.

\bibitem{capprobe}
\BIBentryALTinterwordspacing
R.~Kapoor, L.-J. Chen, L.~Lao, M.~Gerla, and M.~Y. Sanadidi, ``{CapProbe: A
  Simple and Accurate Capacity Estimation Technique},'' in \emph{Proceedings of
  the 2004 Conference on Applications, Technologies, Architectures, and
  Protocols for Computer Communications}, ser. SIGCOMM ’04.\hskip 1em plus
  0.5em minus 0.4em\relax New York, NY, USA: Association for Computing
  Machinery, 2004, p. 67–78. [Online]. Available:
  \url{https://doi.org/10.1145/1015467.1015476}
\BIBentrySTDinterwordspacing

\bibitem{scapy}
\BIBentryALTinterwordspacing
{The Scapy Community}. (2020) {Scapy}. [Online]. Available:
  \url{https://scapy.net/}
\BIBentrySTDinterwordspacing

\bibitem{rfc3954}
\BIBentryALTinterwordspacing
B.~Claise, ``{Cisco Systems NetFlow Services Export Version 9},'' RFC 3954,
  Internet Engineering Task Force, Oct. 2004. [Online]. Available:
  \url{http://www.ietf.org/rfc/rfc3954.txt}
\BIBentrySTDinterwordspacing

\bibitem{rfc7011}
\BIBentryALTinterwordspacing
B.~Claise, B.~Trammell, and P.~Aitken, ``{Specification of the IP Flow
  Information Export (IPFIX) Protocol for the Exchange of Flow Information},''
  RFC 7011, Internet Engineering Task Force, Sep. 2013. [Online]. Available:
  \url{http://www.ietf.org/rfc/rfc7011.txt}
\BIBentrySTDinterwordspacing

\bibitem{rfc1157}
\BIBentryALTinterwordspacing
J.~Case, M.~Fedor, M.~Schoffstall, and J.~Davin, ``{Simple Network Management
  Protocol (SNMP)},'' RFC 1157, Internet Engineering Task Force, May 1990.
  [Online]. Available: \url{http://www.ietf.org/rfc/rfc1157.txt}
\BIBentrySTDinterwordspacing

\bibitem{moloch}
\BIBentryALTinterwordspacing
{Moloch Developers}. (2020) {Moloch Full Packet Capture}. [Online]. Available:
  \url{https://molo.ch/}
\BIBentrySTDinterwordspacing

\bibitem{mrtg}
\BIBentryALTinterwordspacing
{Tobi Oetiker}. (2020) {The Multi Router Traffic Grapher}. [Online]. Available:
  \url{https://oss.oetiker.ch/mrtg/}
\BIBentrySTDinterwordspacing

\bibitem{opennms}
\BIBentryALTinterwordspacing
{The OpenNMS Group}. (2020) {The OpenNMS Platform}. [Online]. Available:
  \url{https://www.opennms.com/}
\BIBentrySTDinterwordspacing

\bibitem{cacti}
\BIBentryALTinterwordspacing
{The Cacti Group}. (2020) {Cacti}. [Online]. Available:
  \url{https://www.cacti.net/}
\BIBentrySTDinterwordspacing

\bibitem{zabbix}
\BIBentryALTinterwordspacing
{Zabbix LLC}. (2020) {Zabix}. [Online]. Available: \url{https://www.zabbix.com}
\BIBentrySTDinterwordspacing

\bibitem{plum-island}
\BIBentryALTinterwordspacing
{Lily Hay Newman}. The hail mary plan to restart a hacked {US} electric grid.
  [Online]. Available:
  \url{https://www.wired.com/story/black-start-power-grid-darpa-plum-island}
\BIBentrySTDinterwordspacing

\bibitem{rfc5795}
\BIBentryALTinterwordspacing
K.~Sandlund, G.~Pelletier, and L.-E. Jonsson, ``{The RObust Header Compression
  (ROHC) Framework},'' RFC 5795, Internet Engineering Task Force, Mar. 2010.
  [Online]. Available: \url{http://www.ietf.org/rfc/rfc5795.txt}
\BIBentrySTDinterwordspacing

\end{thebibliography}

\smallskip

\glsresetall\footnotesize\noindent
This research was developed with funding from the \gls{DARPA}. The views and conclusions contained in this document are those of the authors and should not be interpreted as representing the official policies, either expressed or implied, of the \glsdesc{DARPA} or the U.S. government. Distribution statement A. Distribution approved for public release, distribution unlimited. Not export controlled per ES-FL-020821-0013.

\end{document}